\documentclass[12pt]{article}

\usepackage{epsf,amsfonts,hyperref}
\renewcommand{\appendix}[1]{
    \addtocounter{section}{1}
    \setcounter{equation}{0}
    \renewcommand{\thesection}{\Alph{section}}
    \section*{Appendix \thesection\protect\indent #1}
    \addcontentsline{toc}{section}{Appendix \thesection\ \ \ #1}
}
\newcommand\encadremath[1]{\vbox{\hrule\hbox{\vrule\kern8pt
\vbox{\kern8pt \hbox{$\displaystyle #1$}\kern8pt}
\kern8pt\vrule}\hrule}}
\def\enca#1{\vbox{\hrule\hbox{
\vrule\kern8pt\vbox{\kern8pt \hbox{$\displaystyle #1$}
\kern8pt} \kern8pt\vrule}\hrule}}

\newcommand\figureframex[3]{
\begin{figure}[bth]
\hrule\hbox{\vrule\kern8pt
\vbox{\kern8pt \vbox{
\begin{center}
{\mbox{\epsfxsize=#1.truecm\epsfbox{#2}}}
\end{center}
\caption{#3}
}\kern8pt}
\kern8pt\vrule}\hrule
\end{figure}
}
\newcommand\figureframey[3]{
\begin{figure}[bth]
\hrule\hbox{\vrule\kern8pt
\vbox{\kern8pt \vbox{
\begin{center}
{\mbox{\epsfysize=#1.truecm\epsfbox{#2}}}
\end{center}
\caption{#3}
}\kern8pt}
\kern8pt\vrule}\hrule
\end{figure}
}

\renewcommand{\thesection}{\arabic{section}}

\makeatletter
\@addtoreset{equation}{section}
\makeatother
\newtheorem{theorem}{Theorem}[section]

\newtheorem{remark}{Remark}[section]
\newtheorem{proposition}{Proposition}[section]
\newtheorem{lemma}{Lemma}[section]
\newtheorem{corollary}{Corollary}[section]
\newtheorem{definition}{Definition}[section]
\def\br{\begin{remark}\rm\small}
\def\er{\end{remark}}
\def\bt{\begin{theorem}}
\def\et{\end{theorem}}
\def\bd{\begin{definition}}
\def\ed{\end{definition}}
\def\bp{\begin{proposition}}
\def\ep{\end{proposition}}
\def\bl{\begin{lemma}}
\def\el{\end{lemma}}
\def\bc{\begin{corollary}}
\def\ec{\end{corollary}}
\def\beaq{\begin{eqnarray}}
\def\eeaq{\end{eqnarray}}
\newcommand{\proof}[1]{{\noindent \bf proof:}\par
{#1} $\square$}

\newcommand{\rf}[1]{(\ref{#1})}
\newcommand{\eq}[1]{Eq.~(\ref{#1})}

\newcommand{\beq}{\begin{equation}}
\newcommand{\eeq}{\end{equation}}
\newcommand{\bea}{\begin{eqnarray}}
\newcommand{\eea}{\end{eqnarray}}

\renewcommand{\and}{{\qquad {\rm and} \qquad}}

\newcommand{\virg}{{\qquad , \qquad}}

 \newcommand{\Tr}{{\,\rm Tr}\:}

\newcommand{\Res}{\mathop{\,\rm Res\,}}

\newcommand{\td}[1]{{\tilde{#1}}}

\newcommand{\om}{\omega}

\newcommand{\ee}[1]{{{\rm e}^{#1}}}

\renewcommand{\d}{{{\partial}}}

\newcommand{\R} {{\mathbf{R\,}}}

\newcommand{\Pint}{{\int\kern -1.em -\kern-.25em}}

\renewcommand{\Re}{{\mathrm{Re}}}

\newcommand{\Ecal}{{\cal E}}
\newcommand{\Dcal}{{\cal D}}

\newcommand{\Id}{{\rm Id}\,}

\newcommand{\calW}{{\cal W}}

\newcommand{\tdU}{{\td{U}}}

\newcommand{\wPsi}{{\vec\psi}}
\newcommand{\wPhi}{{\vec\phi}}
\newcommand{\bfp}{{\mathbf{\psi}_{\infty}}}
\newcommand{\bfq}{{\mathbf{\phi}_{\infty}}}
\newcommand{\bfe}{{\mathbf{e}}}

\begin{document}
%=============================Page de titre===============
%\date{??}
%\author{Eynard}
%\title{Correlation functions for hermitian random matrices}
%\topmargin .5cm \textheight 21.5cm \textwidth 15.8cm
%\oddsidemargin 0.54cm
%\evensidemargin 0.54cm
\sloppy

%\maketitle

\pagestyle{empty}
\hfill SPhT-T06/037
\addtolength{\baselineskip}{0.20\baselineskip}
\begin{center}
\vspace{26pt}
{\large \bf {Mixed correlation function and spectral curve for the 2-matrix model}}
\newline
\vspace{26pt}

%{\sl M.\ Bertola}\hspace*{0.05cm}\footnote{ E-mail: eynard@spht.saclay.cea.fr },
{\sl M.\ Berg\`ere}\hspace*{0.05cm}\footnote{ E-mail: bergere@spht.saclay.cea.fr },
{\sl B.\ Eynard}\hspace*{0.05cm}\footnote{ E-mail: eynard@spht.saclay.cea.fr }
%\vspace{6pt}
%Service de Physique Th\'{e}orique de Saclay,\\
%F-91191 Gif-sur-Yvette Cedex, France.\\
\end{center}

\vspace{20pt}
\begin{center}
{\bf Abstract}:

We compute the mixed correlation function in a way which involves only the orthogonal polynomials with degrees close to $n$, (in some sense like the Christoffel Darboux theorem for non-mixed correlation functions). We also derive new representations for the differential systems satisfied by the biorthogonal polynomials, and we find new formulae for the spectral curve.
In particular we prove the conjecture of M. Bertola, claiming that the spectral curve is the same curve which appears in the loop  equations.

\end{center}

\tableofcontents

%-----------------------------ABSTRACT--------------------------------------
%
%Abstract

%\begin{center}

%\end{center}

%\newpage
%\pagestyle{empty}

%\section*{}

%\newpage
\vspace{26pt}
\pagestyle{plain}
\setcounter{page}{1}

%*********************************************************************
%==================== ARTICLE ========================================
%*********************************************************************

\section{Introduction}

Consider a pair of random hermitean matrices $M_1$ and $M_2$, of size $n$, with the probability measure:
\beq\label{matrixmodel}
\ee{-\Tr(V_1(M_1)+V_2(M_2)+M_1 M_2)} \,\, dM_1\,dM_2.
\eeq
This random 2-matrix  model has many applications to physics (in particular in quantum gravity, i.e. statistical physics on a random surface and conformal field theory \cite{Kazakov, ZJDFG}) and mathematics (bi-orthogonal polynomials \cite{Mehta}).
Another important application of the 2-hermitean matrix model comes from the fact that it is the analytical continuation of the complex-matrix model, which describes the Dyson gaz  at $\beta=2$ and is an important model of Laplacian growth \cite{WZ}.
And the complex matrix model plays a crucial role in the AdS/CFT correspondance, in the so-called BMN limit \cite{Kristjansen:2002bb}. In that model, it is important to know how to compute mixed expectation values.

\smallskip

For all applications, one would like to be able to compute various expectation values.
Some expectation values can be written in terms of eigenvalues of $M_1$ and $M_2$, for instance $<\Tr M_1^{k} \Tr M_2^{l}>$, which we call non-mixed
because each trace contains only one type of matrix $M_1$ or $M_2$ but not both.
On the contrary, mixed expectation values are those where $M_1$ and $M_2$ may appear together in the same trace, for instance $<\Tr M_1^{k} M_2^{l}>$.
Mixed expectation values cannot be written in terms of eigenvalues of $M_1$ and $M_2$, and are thus more difficult to compute than non-mixed ones.

Beyond the technical challenge of computing them, mixed expectation values should play an important role in applications to boundary conformal field theory and to the BMN limit of string theory/CFT correspondance.
While many formulae for non mixed expectation values have been known for a long time, in particular in terms of bi-orthogonal polynomials \cite{eynardmehta, uvarov, FyodStra, bergere1, bergere2, bergere3, Akemann},
it is only recently that formulae have  been found for mixed traces.
In particular the following expectation value:
\beq
W_n(x,y)=1+\left< \Tr{1\over x-M_1}{1\over y-M_2}\right>,
\eeq
was first computed in \cite{BEmixed}. The idea was to diagonalize the hermitean matrices $M_1=V X V^\dagger$ and $M_2=VU Y  U^\dagger V^\dagger$ where $U$ and $V$ are unitary matrices and $X$ and $Y$ are diagonal matrices containing the eigenvalues of $M_1$ and $M_2$.
Then:
\beq
W_n(x,y)=1+\left< \Tr{1\over x-X}U{1\over y-Y}U^\dagger\right>,
\eeq
and using the Morozov's formula \cite{morozov, eynmorozov} for unitary integrals of the form $<U_{ij} U^\dagger_{kl}>$, one reexpresses $W_n$ in terms of eigenvalues of $M_1$ and $M_2$ only. Then the integration over  eigenvalues is done with the method of biorthogonal polynomials \cite{Mehtabio}.
The result found in \cite{BEmixed} is thus a $n\times n$ determinant involving recursion coefficients of bi-orthogonal polynomials:
\beq
W_n(x,y)= \mathop{\det}_{n\times n}{\left(\Id_n+\Pi_{n-1} {1\over x-Q}{1\over y-P^t}\Pi_{n-1} \right)},
\eeq
where the matrices $Q$ and $P$ implement the recursion relation (multiplication by $x$ and $y$) of the biorthogonal polynomials, and $\Pi_{n-1}$ is the projection on the polynomials of degree $\leq n-1$. Those notations are explained in more details in section \ref{sectionrecursions}.

The mere existence of such a formula was a progress, but a $n\times n$ determinant is not convenient for practical computations and for taking large $n$ limits.
In the non-mixed case, the Christoffel-Darboux theorem allows one to rewrite expectation values in terms of smaller determinants, whose size does not grow with $n$ \cite{eynardmehta, bergere2, bergere3}.

The purpose of the present article is to find a similar property for $W_n(x,y)$, i.e. write it in terms of determinants whose size is independent of $n$.

As a byproduct of such a rewriting, we are able to find new formulae for the spectral curve of the isomonodromic differential system satisfied by the corresponding biorthogonal polynomials.

\subsection{Plan}

\begin{itemize}

\item In section 2, we recall the definitions of bi-orthogonal polynomials, as well as their Fourier-Laplace transforms, Cauchy transforms, and the third type of solution introduced by \cite{Johnsol, Johnsolmtl}.
We also recall their recursion relations, obtained by multiplication by $x$ or derivation. We define the infinite matrices (finite band) $Q$ and $P$, which encode the recursion relations ($Q$ and $P$ are also Lax matrices).

\item In section 3, we study the inverse of $x-Q$ and $y-P$, in particular, we show that they have distinct right and left inverses. The difference between the right and left inverse, is related to the so-called folding matrix \cite{BEformulaD}, and is found to be a sum of bi-orthogonal polynomials and their various transforms.

\item In section 4, we introduce the kernels \cite{eynardmehta}, which are the building blocks of correlation functions. We show that they satisfy some Christoffel Darboux relations.

\item In section 5, we compute the biorthogonal polynomials, their Cauchy transforms, and the kernels as determinants involving matrices $Q$ and $P$.

\item In section 6, we introduce the notion of windows of consecutive bi-orthogonal polynomials \cite{BEHduality, BEHRH, BEformulaD}, because the recursion relations (matrix $Q$ and $P$) allow to rewrite any bi-orthogonal polynomial $p_m$ as a  linear combination of $p_j$ with $n-d_2\leq j \leq n$, where $d_2+1$ is the degree of the potential entering the weight of orthogonality.

\item In section 7, we prove our main result. We compute the mixed correlation function $W_n(x,y)$ in terms of polynomials in the window, and in terms of the kernels.

\item In section 8, we study some consequences of that formula. In particular, we find new representations of the differential system satisfied by a window, and we compute the spectral curve.
We find several new formulae for the spectral curve, and we prove the conjecture of M. Bertola, which claims that the spectral curve is the same curve which was found from loop equations.

\item In section 9, we discuss the consequences of that formula in terms of tau functions.

\item In section 10, we give some explicit examples of our formulae, namely the gaussian and gaussian elliptical cases.

\item Most of the technical proofs are put in the appendix.

\end{itemize}

\subsection{Main results}

This article is very technical, and the purpose is to give many formulae for effective computations with bi-orthogonal polynomials.
We propose some new formulae in almost every paragraph. Let us mention here the most important ones:
\begin{itemize}

\item In section \ref{secinverse}, we give integral representations of the right ($R$) and left ($L$) inverses of $x-Q$. One of the important relations is \eq{relL-RPsiPhi}:
\beq
L(x)-R(x) = \Psi_\infty(x) \, \hat\Phi_\infty^t(x)
\eeq

\item In section \ref{secdetformulae}, we give determinantal expressions of bi-orthogonal polynomials and their transforms, as well as kernels. For instance the kernel $K_n(x,y)$ is:
\beq
K_{n+1}(x,y) \propto \det_{n\times n}\Big((x-Q)(y-P^t)\Big)
\eeq

\item In section \ref{secmixedfction}, we state one of the main results of this paper, i.e. some formulae for the 2-point mixed correlation function $W_n(x,y)$:
\bea
W_n(x,y) &=& 1+\left<\Tr {1\over x-M_1}\,{1\over y-M_2}\right> \cr
&=& \det_{n\times n} \Big(1+{1\over x-Q}{1\over y-P^t}\Big) \cr
&=& \gamma_n^2\,K_{n+1}(x,y)\,J_{n-1}(x,y)\,\times \cr
&& \quad \det_{d_2+1\times d_2+1}\Big(1-\Pi_n^{n-d_2}(1-{\wPsi_n\wPhi_n^t\over K_{n+1}})U_n(1-{\hat\wPsi_n\hat\wPhi_n^t\over J_{n-1}})\tdU_n^t\Big) \cr
\eea
and theorem \ref{thWnkernels} gives $W_n$ in terms of kernels only.

We also find interesting recursion relations for $W_{n+1}-W_n$.

\item In section \ref{secdifsys}, we compute new representations of the differential system $\Dcal_n(x)=\Psi_n'(x)\Psi_n(x)^{-1}$, and we compute the spectral curve:
\beq
\Ecal_n(x,y) = \td{t} \det(y-\Dcal_n(x))  = -\gamma_n^2\, \det(1-U_n\td{U}_n^t) = \det(y-H^{(ij)}(x,x))
\eeq
The most interesting is that we prove Bertola's conjecture \cite{marcoconj}:
\beq
\Ecal_n(x,y) = (V'_1(x)+y)(V'_2(y)+x)-n+\left<\Tr {V'_1(x)-V'_1(M_1)\over x-M_1}\,{V'_2(y)-V'_2(M_2)\over y-M_2}\right>
\eeq
This conjecture has important consequences in terms of tau functions.
Indeed the Miwa-Jimbo-Ueno approach of isomonodromic tau functions \cite{JMU, JM, UT}, generalized in \cite{bertolatau}, allows to express the tau function in terms of residues of the spectral curve, and this formula is particularly convenient for that purpose. It shows that the tau function is the matrix integral, and it shows that some additional parameters could be added to the model.

\end{itemize}

\section{Definitions and notations about bi-orthogonal polynomials}

This section recalls well known facts about bi-orthogonal polynomials, and stands here just for setting notations and describing known properties.
Notations are similar (although with small differences) to those of \cite{BEHRH, BEformulaD}.

\subsection{Measure and integration paths}

Consider the weight:
\beq\label{defweight}
\om(x,y)=\ee{-(V_1(x)+V_2(y)+xy)},
\eeq
where $V_1$ is a complex polynomial of degree $d_1+1$ and $V_2$ a complex polynomial of degree $d_2+1$:
\beq\label{defV}
V_1(x) = \sum_{k=0}^{d_1+1} t_k x^k
\virg
V_2(y) = \sum_{k=0}^{d_2+1} \td{t}_k y^k.
\eeq
We write the leading coefficients of $V'_1$ and $V'_2$:
\beq
t=(d_1+1)\,t_{d_1+1} \virg \td{t}=(d_2+1)\,\td{t}_{d_2+1}.
\eeq

We choose a basis of $d_1$ contours $\gamma^{(i)}$ with $i=1,\dots, d_1$, going from $\infty$ to $\infty$ in sectors where the integral $\int \ee{-V_1(x)}dx$ is convergent,
and we choose a basis of $d_2$ contours $\td\gamma^{(i)}$ with $i=1,\dots, d_2$, going from $\infty$ to $\infty$ in sectors where the integral $\int \ee{-V_2(y)}dy$ is convergent (see \cite{BEHRH, Marcopaths}).

Then we choose a dual basis of $d_1$ contours $\overline{\gamma}^{(i)}$ with $i=1,\dots, d_1$, going from $\infty$ to $\infty$ in sectors where the integral $\int \ee{+V_1(x)}dx$ is convergent,
and we choose a dual basis of $d_2$ contours $\tilde{\overline{\gamma}}^{(i)}$ with $i=1,\dots, d_2$, going from $\infty$ to $\infty$ in sectors where the integral $\int \ee{+V_2(y)}dy$ is convergent,
such that:
\beq\label{dualcontours}
\gamma^{(i)} \cap {\overline{\gamma}}^{(j)} = \delta_{ij}
\virg
\tilde{{\gamma}}^{(i)} \cap \tilde{\overline{\gamma}}^{(j)} = \delta_{ij}.
\eeq

Then, we choose $d_1 d_2$ numbers $\kappa_{i,j}$ such that at least one of them is  non-vanishing, and we define a path $\Gamma$:
\beq
\Gamma:=\sum_{i=1}^{d_1}\sum_{j=1}^{d_2} \kappa_{i,j} \gamma^{(i)}  \times \td\gamma^{(j)},
\eeq
and we define the following measure on $\Gamma$:
\beq\label{defmeasure}
d\mu(x,y)=\ee{-(V_1(x)+V_2(y)+xy)}\,\,dx\, dy.
\eeq

\br{Generalized path $\Gamma$ and matrix models}\label{remGammamatrixmodel}

We have introduced the generalized integration contours $\Gamma$, because it is the most general contour on which the measure $d\mu$ can be integrated.
It corresponds to a generalization of the hermitean 2-matrix model.
Indeed, hermitean matrices have their eigenvalues on the real axis, and the hermitean 2-matrix model \eq{matrixmodel} corresponds to the case $\Gamma=\R\times \R$.

A generalized path $\Gamma$ can also correspond to a matrix model, with matrices which are not hermitean. It corresponds to normal matrices (i.e. which can be diagonalized by a unitary transformation, but with complex eigenvalues), with pairs of eigenvalues constrained to be on $\Gamma$.
This allows to define an ensemble of matrices which is noted $H_n\times H_n(\Gamma)$, see \cite{eynhabilit} for more details.

In this normal matrix model, it makes sense to compute matrix expectation values, in particular the mixed correlation function $W_n(x,y)=1+\left<\Tr {1\over x-M_1}\,{1\over y-M_2}\right>$.

\er

\subsubsection*{More definitions}

We define (see \cite{Johnsol, Johnsolmtl}):
\beq\label{defh0}
h_0 = \int_\Gamma \,d\mu(x,y)
= \sum_{i=1}^{d_1}\sum_{j=1}^{d_2} \kappa_{i,j} \int_{\gamma^{(i)}} dx\, \int_{\td\gamma^{(j)}}dy\,\,\ee{-(V_1(x)+V_2(y)+xy)}.
\eeq
And:
\beq\label{defg}
g^{(i)}(x):= {\sqrt{h_0}\over 2i\pi}\,\int_{\tilde{\overline{\gamma}}^{(i)}} \ee{xy}\,\ee{V_2(y)}\, dy \qquad \,\, ,\, i=1,\dots, d_2\virg g^{(0)}:=0,
\eeq
\beq\label{deftdg}
\td{g}^{(i)}(y):= {\sqrt{h_0}\over 2i\pi}\,\int_{\overline\gamma^{(i)}} \ee{xy}\,\ee{V_1(x)}\, dx \qquad \,\, ,\, i=1,\dots, d_1\virg \td{g}^{(0)}:=0,
\eeq
which are the independent solutions of the differential equations:
\beq
V'_2(\partial/\partial x)g^{(i)}(x) = -x g^{(i)}(x)
\eeq
\beq
V'_1(\partial/\partial y)\td{g}^{(i)}(y) = -y \td{g}^{(i)}(y).
\eeq

Then we define the following ``concomitents'' \cite{Johnsol, Johnsolmtl}:
\beq
c^{(ij)} = {1\over 2i\pi}\,\int_{\gamma^{(i)}} dx  \int_{\overline\gamma^{(j)}} dx' \,\, {V'_1(x)-V'_1(x')\over x-x'}\,\ee{V_1(x')-V_1(x)}\,\ee{(x'-x)y},
\eeq
\beq
\td{c}^{(ij)} = {1\over 2i\pi}\,\int_{\td\gamma^{(i)}} dy  \int_{\td{\overline{\gamma}}^{(j)}} dy' \,\, {V'_2(y)-V'_2(y')\over y-y'}\,\ee{V_2(y')-V_2(y)}\,\ee{(y'-y)x}.
\eeq
The $c^{(ij)}$'s (resp. $\td{c}^{(ij)}$'s) are independent of $y$ (resp. $x$).
Due to the dual choice of contours \eq{dualcontours}, they are normalized:
\beq\label{normalizedcij}
c^{(ij)} =  \delta_{i,j}
\virg
\td{c}^{(ij)} =  \delta_{i,j}.
\eeq
Indeed, integrating by parts, we can replace both $V'_1(x)$ by $y$ and $V'_1(x')$ by $y$. If $i\neq j$, the contours $\gamma^{(i)}$ and $\overline\gamma^{(j)}$ do not intersect, and the integration by parts gives no boundary term and the result vanishes.
If $i=j$, the two contours intersect and we have a boundary term. A way to compute it, is to write the pole $1/(x-x')$ as the sum of a principal part and $2i\pi \delta(x-x')$.
The principal part is integrated by parts and gives zero as in the $i\neq j$ case, whilst the $\delta$-term corresponds to the boundary term in the integration by parts, and it gives $1$.

This computation was first done by the authors of \cite{Johnsol, Johnsolmtl}, in the comparison of the two Riemann-Hilbert problems derived for biorthogonal polynomials \cite{BEHRH, KML, kapaev}

\subsection{Bi-orthogonal polynomials}

The monic bi-orthogonal polynomials \cite{Mehtabio, Mehta}, (if they exist), are uniquely determined by:
\beq
p_n(x)=x^n+\dots
\virg
q_n(y)=y^n+\dots,
\eeq
and
\beq\label{defbiortho}
\int_\Gamma p_n(x)q_m(y)\,d\mu(x,y) =  h_n\,\delta_{nm}.
\eeq
For given potentials $V_1$ and $V_2$, bi-orthogonal polynomials exist for almost every choice of $\Gamma$ (in fact they don't exist only for an enumerable set of $\Gamma$'s, see \cite{BEHRH}).

\subsubsection{Wave functions}

We define:
\beq\label{defwavefunctions}
\psi_n(x):= {1\over\sqrt{h_n}}\,p_n(x)\,\ee{-V_1(x)}
\virg
\phi_n(y):= {1\over\sqrt{h_n}}\,q_n(y)\,\ee{-V_2(y)}.
\eeq

\subsubsection{Cauchy transforms}

We introduce the Cauchy transforms \cite{BEHduality, BEHRH}:
\beq\label{defCauchyp}
\hat\psi_n(y) := {1\over\sqrt{h_n}}\,\ee{V_2(y)}\,\int_\Gamma {1\over y-y'}\, p_n(x') \,d\mu(x',y'),
\eeq
\beq\label{defCauchyq}
\hat\phi_n(x) := {1\over\sqrt{h_n}}\,\ee{V_1(x)}\,\int_\Gamma {1\over x-x'}\, q_n(y') \,d\mu(x',y').
\eeq

\subsubsection{Fourier-Laplace transforms}

We also introduce the following functions \cite{eynardmehta, eynardchain}:
\beq
\hat\psi_n^{(i)}(y) := \int_{\gamma^{(i)}} \psi_n(x) \,\ee{-xy}\, dx
\qquad {\rm for}\,\, i=1,\dots,d_1,
\eeq
\beq
\hat\phi_n^{(i)}(x) := \int_{\td\gamma^{(i)}} \phi_n(y) \,\ee{-xy}\, dy
\qquad {\rm for}\,\, i=1,\dots,d_2.
\eeq
They are the Fourier-Laplace transforms of $\psi_n$ and $\phi_n$.
For $i=0$ we will also write:
\beq
\hat\psi_n^{(0)}(y) := \hat\psi_n(y)
\virg
\hat\phi_n^{(0)}(x) := \hat\phi_n(x).
\eeq

\subsubsection{Third-type functions}

The authors of \cite{Johnsol, Johnsolmtl} have introduced the following functions:
\beq
\phi_n^{(i)}(y) := {1\over 2i\pi}\,{1\over\sqrt{h_n}}\,\int_{\overline\gamma^{(i)}} dx \int_\Gamma d\mu(x',y')\, {1\over y-y'} {V'_1(x)-V'_1(x')\over x-x'} \ee{V_1(x)} \ee{xy} \, q_n(y'),
\eeq
\beq
\psi_n^{(i)}(x) := {1\over 2i\pi}\,{1\over\sqrt{h_n}}\,\int_{\tilde{\overline{\gamma}}^{(i)}} dy \int_\Gamma d\mu(x',y')\, {1\over x-x'} {V'_2(y)-V'_2(y')\over y-y'} \ee{V_2(y)} \ee{xy} \, p_n(x').
\eeq

We will also write:
\beq
\psi_n^{(0)}(x) := \psi_n(x)
\virg
\phi_n^{(0)}(y) := \phi_n(y).
\eeq

Formaly, those functions are the ``inverse Fourier transforms'' of the $\hat\psi$'s, as described in \cite{BEHRH, kapaev}.

\subsection{Semi-infinite vectors and matrices}

We introduce semi-infinite vector notations:
\beq\label{defpvec}
\bfp(x) = (\psi_0(x),\psi_1(x),\psi_2(x),\dots)^t
\virg
\bfq(y) = (\phi_0(y),\phi_1(y),\phi_2(y),\dots)^t,
\eeq
and more generaly:
\beq\label{defpveci}
\bfp^{(i)}(x) = (\psi^{(i)}_0(x),\psi^{(i)}_1(x),\psi^{(i)}_2(x),\dots)^t
\,\, , \,\,
\bfq^{(i)}(y) = (\phi^{(i)}_0(y),\phi^{(i)}_1(y),\phi^{(i)}_2(y),\dots)^t.
\eeq
And
\beq\label{defphatvec}
\hat\bfp(y) = (\hat\psi_0(y),\hat\psi_1(y),\hat\psi_2(y),\dots)^t
\virg
\hat\bfq(x) = (\hat\phi_0(x),\hat\phi_1(x),\hat\phi_2(x),\dots)^t,
\eeq
and more generaly:
\beq\label{defphatveci}
\hat\bfp^{(i)}(y) = (\hat\psi^{(i)}_0(y),\hat\psi^{(i)}_1(y),\hat\psi^{(i)}_2(y),\dots)^t
\,\, , \,\,
\hat\bfq^{(i)}(x) = (\hat\phi^{(i)}_0(x),\hat\phi^{(i)}_1(x),\hat\phi^{(i)}_2(x),\dots)^t.
\eeq

We also introduce the basis vectors:
\beq\label{defevec}
\bfe_n=(\,\mathop{\overbrace{0,\dots,0}}^{n},1,\mathop{\overbrace{0,\dots}}^{\infty}\,)^t,
\eeq
i.e. the vector whose only non vanishing component is in the $n+1$ position.
It is such that:
\beq\label{evecpvecpn}
\bfe_n^t\, \bfp(x) =  \psi_n(x).
\eeq

Similarly, we consider the projection matrix:
\beq\label{defPin}
\Pi_n = {\rm diag}(\,\mathop{\overbrace{1,\dots,1}}^{n+1},\mathop{\overbrace{0,\dots}}^{\infty}\,) = \sum_{j=0}^n \bfe_j \bfe_j^t,
\eeq
with $n+1$ ones on the diagonal. It is the projector on the span of $\bfe_0,\bfe_1,\dots,\bfe_n$.
We also define:
\beq
\Pi^n = 1-\Pi_{n-1} = {\rm projector \, on \, the \, span\, of \,}n,n+1,\dots,\infty,
\eeq
and:
\beq
\Pi^n_m = \sum_{j=n}^m \bfe_j \bfe_j^t = \Pi^n\Pi_m = \Pi_m - \Pi_{n-1}  = {\rm projector \, on \, the \, span\, of \,}n,\dots,m.
\eeq

We also introduce the following $\infty\times (d_2+1)$, (resp. $\infty\times (d_1+1)$) matrices:
\beq
\Psi_\infty := \pmatrix{\bfp^{(0)} & \bfp^{(1)} & \dots & \bfp^{(d_2)}},
\eeq
\beq
\Phi_\infty := \pmatrix{\bfq^{(0)} & \bfq^{(1)} & \dots & \bfq^{(d_1)}},
\eeq
\beq
\hat\Psi_\infty := \pmatrix{\hat\bfp^{(0)} & \hat\bfp^{(1)} & \dots & \hat\bfp^{(d_1)}},
\eeq
\beq
\hat\Phi_\infty := \pmatrix{\hat\bfq^{(0)} & \hat\bfq^{(1)} & \dots & \hat\bfq^{(d_2)}}.
\eeq

\subsection{Recursion relations for the bi-orthogonal polynomials}
\label{sectionrecursions}

It is well known\footnote{$x$ and $\d_x$ acting on a polynomial, gives a polynomial which can be decomposed on the basis of biorthogonal polynomials.
}  that we have the following recursion relations \cite{ZJDFG, BEHduality}:
\beq\label{xQrecp}
x\psi_n=\sum_m Q_{nm} \psi_m
\virg
y\phi_n=\sum_m P_{nm} \phi_m,
\eeq
\beq\label{dxpn}
\psi'_n = \sum_m P_{mn} \psi_m
\virg
\phi'_n = \sum_m Q_{mn} \phi_m.
\eeq
Since $V_1$ and $V_2$ are polynomials, $Q$ and $P$ must be finite-band matrices, i.e.:
\beq\label{Qfiniteband}
Q_{nm}\neq0 \quad {\rm iff}\,\, n-d_2\leq m\leq n+1,
\eeq
\beq\label{Pfiniteband}
P_{nm}\neq0 \quad {\rm iff}\,\, n-d_1\leq m\leq n+1.
\eeq

In vector notations we have:
\beq\label{xQrecvecp}
x\bfp = Q\bfp
\virg
y\bfq = P\bfq,
\eeq
\beq\label{dxPrecvecp}
\psi'_\infty = P^t \bfp
\virg
\phi'_\infty = Q^t \bfq.
\eeq

\subsection{Relations between Q and P}

The matrices $Q$ and $P$ have the following properties \cite{ZJDFG, BEHduality, BEHRH}:
\beq\label{PQcannonical}
[Q,P^t]=\Id.
\eeq

\beq\label{defgamma}
Q_{n,n+1}=P_{n,n+1}=\sqrt{h_{n+1}\over h_n}:=\gamma_{n+1}.
\eeq

$P^t+V'_1(Q)$ is a strictly lower triangular matrix
\beq\label{PV'_1+}
(P^t+V'_1(Q))_+=0
\virg
(P^t+V'_1(Q))_{n,n-1}= {n\over \gamma_n},
\eeq
and $Q^t+V'_2(P)$ is a strictly lower triangular matrix
\beq\label{QV'_2+}
(Q^t+V'_2(P))_+=0
\virg
(Q^t+V'_2(P))_{n,n-1}= {n\over \gamma_n}.
\eeq

\subsection{Recursion relations for the Cauchy transforms}

The Cauchy transforms also satisfy recursion relations \cite{BEHRH}:
\beq\label{xQrechatvecp}
y\hat\bfp(y) = P^t\hat\bfp(y) + {1\over \phi_0(y)}\,\bfe_0
\virg
x\hat\bfq(x) = Q^t\hat\bfq(x) + {1\over \psi_0(x)}\,\bfe_0,
\eeq
\beq\label{dxPrechatvecp}
\hat\psi'_\infty = -Q\hat\bfp + {1\over \phi_0}\,{V'_2(y)-V'_2(P^t)\over y-P^t}\bfe_0,
\eeq
\beq\label{dxPrechatvecq}
\hat\phi'_\infty = -P \hat\bfq + {1\over \psi_0}\,{V'_1(x)-V'_1(Q^t)\over x-Q^t} \bfe_0.
\eeq

\subsection{Recursion relations for the Fourier-Laplace transforms}

For $i\neq 0$ we have \cite{BEHduality, BEHRH}:
\beq\label{recxFourrier}
y\hat\bfp^{(i)} = P^t\,\hat\bfp^{(i)}
\virg
x\hat\bfq^{(i)} = Q^t\,\hat\bfq^{(i)},
\eeq

\beq
\hat\psi_\infty^{'(i)} = - Q\,\hat\bfp^{(i)}
\virg
\hat\phi^{'(i)}_\infty = - P\,\hat\bfq^{(i)}.
\eeq
Notice that they satisfy the same recursion relation as the Cauchy transforms without the non-homogeneous term.

\subsection{Recursion relations for the third type functions}

For $i\neq 0$ we have \cite{BEHRH, Johnsol, Johnsolmtl}:
\bea
x\bfp^{(i)}
&=& Q\,\bfp^{(i)} +  \left(V'_2(\partial/\partial x) - V'_2(P^t)\over \partial/\partial x - P^t\right) \bfe_0 \,g^{(i)}  \cr
&=& Q\,\bfp^{(i)} +  {\sqrt{h_0}\over 2i\pi}\,\int_{\td{\overline\gamma}^{(i)}} dy\, \ee{V_2(y)}\,\ee{xy}\, \left(V'_2(y) - V'_2(P^t)\over y - P^t\right) \bfe_0,
\eea
\beq
\psi_\infty^{'(i)} = P^t \,\bfp^{(i)} - g^{(i)}\,\bfe_0,
\eeq

\beq\label{yPrecphii}
y\bfq^{(i)} = P\,\bfq^{(i)} +  \left(V'_1(\partial/\partial y) - V'_1(Q^t)\over \partial/\partial y - Q^t\right) \bfe_0 \,\td{g}^{(i)},
\eeq
\beq
\phi_\infty^{'(i)} = Q^t \,\bfq^{(i)} - \td{g}^{(i)}\,\bfe_0.
\eeq
Notice that they satisfy the same recursion relation as the wave functions, with an additional non-homogeneous term.

\section{Inverses}

The formula for mixed correlation functions found in \cite{BEmixed}, is written in terms of the inverse operators of $x-Q$ and $y-P$,
thus we study them in detail in this section.
$x-Q$ and $y-P$ also have distinct right and left inverses, which were shown to play a crucial role in the notion of folding onto a window in \cite{BEformulaD}.

\bigskip

\subsection{Inverse}
\label{secinverse}

By definition, the infinite matrix $1/(x-Q)$ has elements:
\bea
\left({1\over x-Q}\right)_{nm}
&=& {1\over \sqrt{h_n\,h_m}}\,\int_\Gamma q_m(y')\,{1\over x-x'}\,p_n(x')\,d\mu(x',y') \cr
&=& \psi_n(x)\hat\phi_m(x) +R_{nm}(x),
\eea
where
\bea
R_{nm}(x):={-1\over \sqrt{h_n\,h_m}}\,\int_\Gamma q_m(y')\,{p_n(x)-p_n(x')\over x-x'}\,d\mu(x',y')
\eea
which is a polynomial in $x$ of degree $n-1$.

In vector notations we have:
\beq\label{inverseRpqhat}
{1\over x-Q} = \bfp(x) \hat\bfq^t(x) + R(x).
\eeq
Similarly
\beq
{1\over y-P} = \bfq(y) \hat\bfp^t(y) + \td{R}(y)
\eeq
where
\bea
\td{R}_{mn}(y):=-{1\over \sqrt{h_n\,h_m}}\,\int_\Gamma {q_n(y)-q_n(y')\over y-y'}\,p_m(x')\,d\mu(x',y').
\eea

\subsection{Right inverse}

The semi-infinite matrix $R(x)$ (resp. $\td{R}(y)$) is polynomial in $x$ (resp. $y$) and is strictly lower triangular:
\beq
R_{nm}=0 \quad {\rm if} \,\, m\geq n
\qquad
({\rm resp.}\,\,
\td{R}_{nm}=0 \quad {\rm if} \,\, m\geq n).
\eeq
It is a right inverse \cite{BEformulaD} of $(x-Q)$ (resp. $(y-P)$):
\beq
(x-Q)R(x)=\Id
\qquad
({\rm resp.}\,\,
(y-P)\td{R}(y)=\Id).
\eeq
But it is not a left inverse, we have:
\beq
R(x)(x-Q) = \Id - {\bfp(x)\,\bfe_0^t\over \psi_0(x)}
\qquad
({\rm resp.}\,\,
\td{R}(y)(y-P) = \Id -  {\bfq(y)\,\bfe_0^t\over \phi_0(y)}).
\eeq

Notice that
\beq
R_{n,n-1} = -{1\over \gamma_n} = \td{R}_{n,n-1}.
\eeq

Notice that we have:
\beq
{\psi_n(x)-\psi_n(x')\over x-x'} = - \sum_m R_{nm}(x)\, \psi_m(x')
\eeq
\beq
({\rm resp.}\,\,
{\phi_n(y)-\phi_n(y')\over y-y'} = - \sum_m \td{R}_{nm}(y)\, \phi_m(y')).
\eeq
In particular at $x=x'$:
\beq
\psi'_\infty(x) = -R(x)\bfp(x) = P^t \bfp(x),
\eeq
\beq\qquad
({\rm resp.}\,\,
\phi'_\infty(y) = -\td{R}(y)\bfq(y) = Q^t \bfq(y)).
\eeq

Notice also that:
\beq
R(x_1)\,R(x_2) =-{R(x_1)-R(x_2)\over x_1-x_2}
\eeq
and
\beq
[R(x_1),R(x_2)]=0.
\eeq

It can be found, by solving directely the system $(x-Q)R(x)=1$ for a lower triangular matrix $R$, that:
\beq
R_{nm}(x) = \left\{\begin{array}{ll}
{-1\over \gamma_{m+1}\dots\gamma_{n}}\,\det\pmatrix{\Pi_{m+1}^{n-1}(x-Q)\Pi_{m+1}^{n-1}} & \quad \hbox{for }\,m\leq n-2 \cr
{-1\over \gamma_{n}} & \quad \hbox{for }\,m=n-1 \cr
0 & \quad \hbox{for }\,m\geq n
\end{array}\right.
\eeq

\subsection{Left inverse}

\begin{figure}[bth]
\hrule\hbox{\vrule\kern8pt
\vbox{\kern8pt \vbox{
\begin{center}
{\mbox{\epsfysize=6.truecm\epsfbox{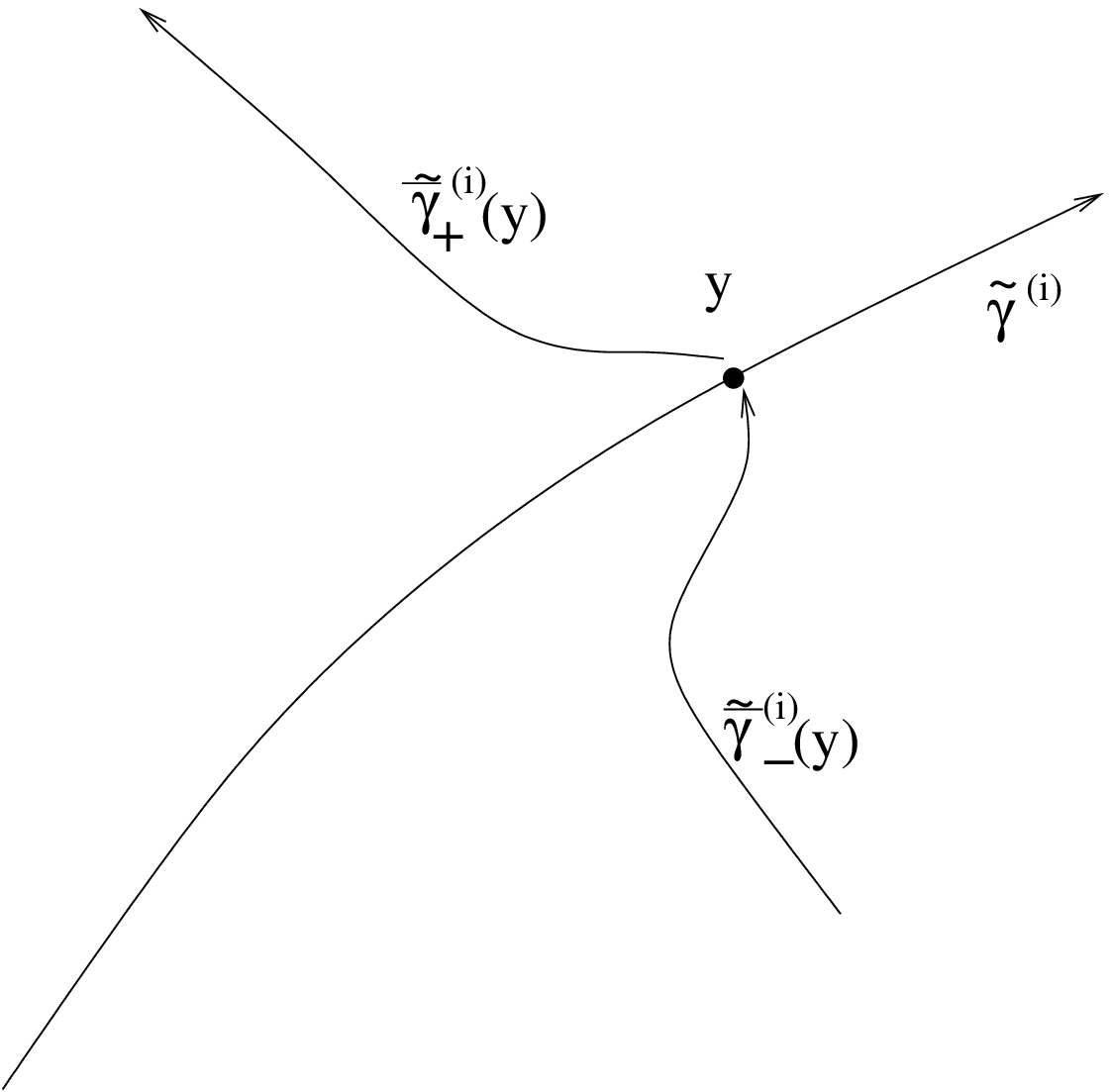}}}
\end{center}
\caption{Definition of the contours $\td{\overline\gamma}^{(i)}_+(y)$ and $\td{\overline\gamma}^{(i)}_-(y)$.\label{defcontoursypm}}
}\kern8pt}
\kern8pt\vrule}\hrule
\end{figure}

Consider $y\in\td\gamma^{(i)}$, then deform the contour $\td{\overline\gamma}^{(i)}$ so that it crosses $\td{\gamma}^{(i)}$at $y$, and define $\td{\overline\gamma}^{(i)}_-(y)$ the part of $\td{\overline\gamma}^{(i)}$ which stands on the right of $\td{\gamma}^{(i)}$ and which ends at $y$, and $\td{\overline\gamma}^{(i)}_+(y)$ the part of $\td{\overline\gamma}^{(i)}$ which stands on the left of $\td{\gamma}^{(i)}$ and which starts at $y$ (see fig.\ref{defcontoursypm}). Then define:
\beq
L_{nm}(x):= {-1\over 2i\pi}\,\sum_{i=1}^{d_2}\,\int_{\td{\gamma}^{(i)}} dy \phi_m(y)\,\ee{-xy}\,\left(\int_{\td{\overline\gamma}^{(j)}_+(y)} dy'' +\int_{\td{\overline\gamma}^{(j)}_-(y)} dy'' \right) \hat\psi_n(y'')\,\ee{xy''}
\eeq
It has the following properties:
\bt\label{thleftinv}
$L_{nm}(x)$ is a polynomial in $x$.

$L$ is upper triangular, with $L_{nm}(x)=0$ if $n+d_2> m$.

$L$ is a left inverse of $x-Q$:
\beq
L(x)(x-Q)=1
\eeq

We have:
\beq
L_{nm}(x) = \left({1\over x-Q}\right)_{nm} + \sum_{i=1}^{d_2}  \psi_n^{(i)}(x) \hat\phi_m^{(i)}(x)
\eeq
\et

\proof{The proof is in appendix A}

\subsubsection*{Alternative definition 1}

It can be found, by solving directely the system $L(x)(x-Q)=1$ for an upper triangular matrix $L$, that:
\beq
L_{nm}(x) = \left\{\begin{array}{ll}
{-1\over Q_{n+d_2,n}\dots Q_{m,m-d_2}}\,\det\pmatrix{\Pi_{n+d_2}^{m-1}(x-Q)\Pi_{n+1}^{m-d_2}} & \quad \hbox{for }\,m>n+d_2 \cr
{-1\over Q_{n+d_2,n}} & \quad \hbox{for }\,m=n+d_2 \cr
0 & \quad \hbox{for }\,m< n+d_2 \cr
\end{array}\right.
\eeq
This proves that there is only one left inverse of $x-Q$ which is upper triangular.

\subsubsection*{Alternative definition 2}

Define:
\beq
l(x):=\sum_i{\td{R}^t(y_i(x))\over V''_2(y_i(x))}
\eeq
where $y_i(x)$ are the $d_2$ solutions of $V'_2(y)+x=0$.
Notice that $l(x)$ is an upper triangular matrix such that:
\beq
l_{nm}(x)=0 \quad {\rm if}\,\,\, n+d_2> m.
\eeq

The left inverse \cite{BEformulaD} of $x-Q$ is given by:
\bea
L(x)&:=&-l(x)\,\, {1\over 1+(V'_2(P^t)+Q) l(x) } \cr
&=& -l(x) + l(x)(V'_2(P^t)+Q) l(x) \cr
&& \qquad \quad - l(x)(V'_2(P^t)+Q) l(x)(V'_2(P^t)+Q) l(x) + \dots \, . \cr
\eea
Since $l(x)$ and $V'_2(P^t)+Q$ are strictly upper triangular matrices, each entry of that infinite sum, is actualy a finite sum.
$L(x)$ is a strictly upper triangular matrix, such that:
\beq
L_{nm}(x)=0 \quad {\rm if}\,\,\, n+d_2> m.
\eeq
This definition gives clearly a left inverse of $x-Q$, and since the upper triangular left inverse of $x-Q$ is unique, it must coincide with the first definition.

\medskip

Remark: $l$ computes the Euclidean division of $q_m$ by $V'_2(y)+x$:
\beq
\phi_m(y) = -(V'_2(y)+x) \sum_n l_{nm}(x)\phi_n(y) + (\deg\leq d_2-1).
\eeq

Similarly, we define:
\beq
\td{l}(y):=\sum_i{R^t(x_i(y))\over V''_1(x_i(y))},
\eeq
and the left inverse of $y-P$ is:
\bea
\td{L}(y)&:=&-\td{l}(y)\,\, {1\over 1+(V'_1(Q^t)+P) \td{l}(y) } \cr
&=& -\td{l}(y) + \td{l}(y)(V'_1(Q^t)+P) \td{l}(y)  \cr
&& \qquad \quad - \td{l}(y)(V'_1(Q^t)+P) \td{l}(y)(V'_1(Q^t)+P) \td{l}(y) + \dots \, .\cr
\eea

\subsection{Relationship between right and left inverses}

We have:
\bea
\left({1\over x-Q}\right)_{nm}
&=&  R_{nm}(x) + \psi_n^{(0)}(x) \hat\phi^{(0)}_m(x) \cr
&=& L_{nm}(x) - \sum_{i=1}^{d_2} \psi_n^{(i)}(x) \hat\phi^{(i)}_m(x)
\eea
i.e. in vector notations
\beq
{1\over x-Q} = L(x) - \sum_{i=1}^{d_2} \bfp^{(i)}(x) \hat\bfq^{(i)t}(x) = R(x) + \bfp(x)\hat\bfq^t(x).
\eeq

This implies:
\beq\label{relL-RPsiPhi}
L(x)-R(x) = \sum_{i=0}^{d_2}  \bfp^{(i)}(x) \hat\bfq^{(i)t}(x) = \Psi_\infty \, \hat\Phi_\infty^t
\eeq
which is a matrix of rank $d_2+1$.

Since $R$ is lower triangular, we have, for $n\leq m$:
\beq
L_{nm}(x) = \sum_{i=0}^{d_2}  \psi_n^{(i)}(x) \hat\phi_m^{(i)}(x) \qquad {\rm if}\, n\leq m,
\eeq
and since $L$ is upper triangular, we have for $n\geq m-d_2+1$:
\beq
R_{nm}(x) = - \sum_{i=0}^{d_2}  \psi_n^{(i)}(x) \hat\phi_m^{(i)}(x) \qquad {\rm if}\, n\geq m-d_2+1.
\eeq
Notice that if $n\leq m \leq n+d_2-1$, we have:
\beq
 \sum_{i=0}^{d_2}  \psi_n^{(i)}(x) \hat\phi_m^{(i)}(x) = 0 \qquad {\rm if}\, n\leq m\leq n+d_2-1,
\eeq
and thus:
\beq
\left({1\over x-Q}\right)_{nm} = \left\{\begin{array}{ll}
\psi_n(x) \hat\phi_m(x)  & \qquad {\rm if}\, n\leq m \cr
- \sum_{i=1}^{d_2}  \psi_n^{(i)}(x) \hat\phi_m^{(i)}(x) & \qquad {\rm if}\, n\geq m-d_2+1 \cr
\end{array}\right.
\eeq

\section{Kernels}

The kernels are the building blocks of non-mixed expectation values \cite{eynardmehta, bergere1}, and we need them because they will appear in mixed expectation values as well.
Thus we recall their definitions and some properties below.

\subsection{Definition of the kernels}

\subsubsection{$K$ kernels}

We introduce the kernel:
\beq\label{defK}
K_n(x,y):=\sum_{m=0}^{n-1} \phi_m(y)\, \psi_m(x) =  \bfq^t(y)\,\Pi_{n-1}\,\bfp(x)
\eeq
and more generaly:
\bea\label{defKij}
K_n^{(i,j)}(x,y)
&:=& \sum_{m=0}^{n-1} \phi^{(i)}_m(y)\, \psi^{(j)}_m(x) + K_0^{(i,j)}(x,y) \cr
&=&  \bfq^{(i)t}(y)\,\Pi_{n-1}\,\bfp^{(j)}(x) + K_0^{(i,j)}(x,y)
\eea
where
\beq
K_0^{(0,0)}:= 0
\eeq
\beq
K_0^{(i,0)}(x,y) := -{1\over 2i\pi}\,\ee{-V_1(x)}\,\int_{\overline\gamma^{(i)}} {1\over x-x'}\, \ee{V_1(x')}\,\ee{x' y} \,dx'
\eeq
\beq\label{defK0i0}
K_0^{(0,j)}(x,y) := -{1\over 2i\pi}\,\ee{-V_2(y)}\,\int_{\td{\overline\gamma}^{(j)}} {1\over y-y'}\, \ee{V_2(y')}\,\ee{x y'}\, dy'
\eeq
\bea\label{defK0ij}
K_0^{(i,j)}(x,y)
& :=&  {1\over (2i\pi)^2}\,\int_{\td{\overline\gamma}^{(i)}}dx'' \int_{\overline\gamma^{(j)}}dy'' \int_\Gamma d\mu(x',y') \ee{V_1(x'')}\ee{V_2(y'')}\,\ee{xy''+x''y}\cr
 && \left({1\over (x-x')(x-x'')}\,{V'_2(y')-V'_2(y'')\over y'-y''}+{1\over (y-y')(y-y'')}\,{V'_1(x')-V'_1(x'')\over x'-x''}\right)  \cr
 &&
\eea

\subsubsection{$J$ kernels}

We also introduce the kernels:
\beq\label{defH}
J_n(x,y):= J_0(x,y)-\sum_{m=0}^{n-1} \hat\phi_m(x) \hat\psi_m(y)
:= J_0(x,y) - \hat\bfq(x)^t \Pi_{n-1} \hat\bfp(y)
\eeq
and more generaly
\beq\label{defHij}
J^{(i,j)}_n(x,y):= J^{(i,j)}_0(x,y)-\sum_{m=0}^{n-1} \hat\phi^{(i)}_m(x) \hat\psi^{(j)}_m(y)
\eeq
where
\beq\label{defH0}
J_0(x,y):=J_0^{(0,0)}(x,y) := \ee{V_1(x)}\,\ee{V_2(y)}\,\int_\Gamma {1\over x-x'}\,{1\over y-y'}\, \,d\mu(x',y')
\eeq
\beq\label{defH00i}
J_0^{(0,i)}(x,y) := \ee{V_1(x)}\,\int_{\gamma^{(i)}} {1\over x-x'}\, \ee{-V_1(x')}\,\ee{-x' y} \,dx'
\eeq
\beq\label{defH0i0}
J_0^{(i,0)}(x,y) := \ee{V_2(y)}\,\int_{\td\gamma^{(i)}} {1\over y-y'}\, \ee{-V_2(y')}\,\ee{-x y'}\, dy'
\eeq
\beq\label{defH0ij}
J_0^{(i,j)}(x,y) := \ee{-x y}
\eeq

Sometimes one writes abusively: $J_n(x,y)= \sum_{m=n}^{\infty} \hat\phi_m(x) \hat\psi_m(y)=\hat\bfq^t \Pi^n \hat\bfp$.

\subsubsection{$H$ kernels}

We also introduce the kernels:
\beq\label{defhatK}
H^{(ij)}_n(x_1,x_2) := \sum_{m=0}^{n-1} \hat\phi^{(i)}_m(x_1)\,\psi^{(j)}_m(x_2) + H^{(ij)}_0(x_1,x_2)
\eeq

where
\beq
H_0^{(i,j)}(x_1,x_2) :=  {1\over x_2-x_1}\, {\ee{V_1(x_1)-V_1(x_2)}\over 2i\pi}\,\int_{\td{\overline\gamma}^{(j)}}dy \int_{\td{\gamma}^{(i)}} dy' \,{V'_2(y)-V'_2(y')\over y-y'}\, {\om(x_1,y')\over \om(x_2,y)}
\eeq
\beq
H_0^{(0,j)}(x_1,x_2) :=  {\ee{V_1(x_1)-V_1(x_2)}\over 2i\pi}\,\int_{\td{\overline\gamma}^{(j)}}dy \int_\Gamma dx' dy' \,{V'_2(y)-V'_2(y')\over y-y'}\,{1\over x'-x_1}\,{1\over x'-x_2}\, {\om(x',y')\over \om(x_2,y)}
\eeq
\beq
H_0^{(i,0)}(x_1,x_2) :=0
\eeq
\beq
H_0^{(0,0)}(x_1,x_2) := {1\over x_2-x_1}\,\ee{V_1(x_1)-V_1(x_2)}
\eeq

Notice that at $x_1=x_2$ we have:
\beq\label{Hijx=x}
\mathop{{\rm lim\,}}_{x_1\to x_2}\, (x_2-x_1)H_n^{(i,j)}(x_1,x_2) =  \delta_{i,j}
\eeq

\subsubsection{$\td{H}$ kernels}

Similarly:
\beq\label{defhattdK}
\td{H}^{(ij)}_n(y,y') := \sum_{m=0}^{n-1} q_m^{(i)}(y)\hat{p}^{(j)}_m(y') + \td{H}^{(ij)}_0(y,y')
\eeq

\subsection{Christoffel--Darboux matrices}

We define the Christoffel-Darboux matrices \cite{BEHduality}:
\beq
A_n:=[\Pi_{n-1},Q] = \gamma_n \bfe_{n-1}\bfe_n^t - (1-\Pi_{n-1}) Q \Pi_{n-1},
\eeq
\beq
B_n:=[\Pi_{n-1},P] = \gamma_n \bfe_{n-1}\bfe_n^t - (1-\Pi_{n-1}) P \Pi_{n-1}.
\eeq

Notice that since $Q$ (resp. $P$) is finite band, $A_n$ (resp. $B_n$) is non-vanishing only in a sub--block of size $d_2+1\times d_2+1$ (resp. $d_1+1\times d_1+1$), i.e.
\beq
\left(A_n\right)_{ij}\neq0 \quad {\rm iff}\,\, n-d_2\leq i\leq n\,\,{\rm and}\,\, n-1\leq j \leq n+d_2-1
\eeq
\beq
\left(B_n\right)_{ij}\neq0 \quad {\rm iff}\,\, n-d_1\leq i\leq n\,\,{\rm and}\,\, n-1\leq j \leq n+d_1-1
\eeq
We say that $A_n$ and $B_n$ are {\bf ``small matrices''}, i.e. their size is not growing with $n$.

\subsection{Christoffel--Darboux theorems}
\label{secCDth}

Using the recursion relations \eq{xQrecp} to \eq{yPrecphii},
we have the following Christoffel--Darboux theorems \cite{BEHduality, BEHRH}:

\bt
\beq\label{CDthKhat}
(x_2-x_1) H_n^{(ij)}(x_1,x_2) = \hat\bfq^{(i)\,t}(x_1) A_n \bfp^{(j)}(x_2)
\eeq
\beq
(\partial_{x_1}+\partial_{x_2}) H_n^{(ij)}(x_1,x_2) = - \hat\bfq^{(i)\,t}(x_1) B_n^t \bfp^{(j)}(x_2)
\eeq
\beq
(x-\partial/\partial y) K^{(ij)}_n(x,y) = \bfq^{(i)\,t}(y) A_n \bfp^{(j)}(x)
\eeq
\beq
(y-\partial/\partial x) K^{(ij)}_n(x,y) = \bfq^{(i)\,t}(y) B_n^t \bfp^{(j)}(x)
\eeq
\beq
(x+\partial/\partial y) J_n(x,y) = \hat\bfq^{(i)\,t}(y) A_n \hat\bfp^{(j)}(x)
\eeq
\beq
(y+\partial/\partial x) J_n(x,y) = \hat\bfq^{(i)\,t}(y) B_n^t \hat\bfp^{(j)}(x)
\eeq
\et

In particular, at $x_1=x_2=x$ in \eq{CDthKhat}, and using \eq{Hijx=x}, we recover the duality of \cite{BEHduality}:
\beq\label{hatKx1egalx2}
\hat\bfq^{(i)\,t}(x) A_n \bfp^{(j)}(x) = \delta_{i,j}
\eeq
and we find:
\beq
H_n^{(ij)}(x,x) = \hat\bfq^{(i)\,t}(x) A_n \psi_\infty^{'(j)}(x).
\eeq

\section{Determinantal formula for the orthogonal polynomials and kernels}
\label{secdetformulae}

\subsection{Determinantal formulae}
\label{detformula}

The following formulae are very useful to express kernels or polynomials, as  well as the Cauchy transforms, as $n\times n$ determinants.
The proofs are given in appendix B.

\begin{enumerate}

\item
$\bullet$ determinant of $\Pi_{n-1}(x-Q)\Pi_{n-1}$:
\beq\label{detPix-QPi}
\det(\Pi_{n-1}(x-Q)\Pi_{n-1}) =  p_n(x)
\virg
\det(\Pi_{n-1}(y-P)\Pi_{n-1}) =  q_n(y)
\eeq

\item
$\bullet$ determinant of $\Pi_{n-1}{1\over x-Q}\Pi_{n-1}$:

\beq\label{detPiinvx-QPi}
\det(\Pi_{n-1}{1\over x-Q}\Pi_{n-1}) = {\ee{-V_1(x)}\over \sqrt{h_{n-1}}}\,\hat{\phi}_{n-1}(x)
\eeq
\beq
\det(\Pi_{n-1}{1\over y-P}\Pi_{n-1}) = {\ee{-V_2(y)}\over \sqrt{h_{n-1}}}\,\hat{\psi}_{n-1}(y)
\eeq

\item
$\bullet$ kernel $K_n$:
\beq\label{detPix-Qy-PPi}
\det(\Pi_{n-1} (x-Q)(y-P^t) \Pi_{n-1}) = h_n \, \ee{V_1(x)+V_2(y)}\, K_{n+1}(x,y)
\eeq

\item
$\bullet$ kernel $J_n$:

\beq\label{detPiinvx-Qinvy-PPi}
\det(\Pi_{n-1}{1\over x-Q}{1\over y-P^t}\Pi_{n-1}) = {\ee{-V_1(x)-V_2(y)}\over h_{n-1}}\, J_{n-1}(x,y)
\eeq

\end{enumerate}

\subsection{Inverses}

It is usefull to compute also the inverses (i.e. all minors) of the previous matrices.
The proofs are given in appendix \ref{appproofdetformula}.
The following formula give inverses of $n\times n$ matrices.

\begin{enumerate}

\item
$\bullet$ inverse of $\Pi_{n-1}(x-Q)\Pi_{n-1}$:
\beq\label{inversePix-QPi}
\left(\Pi_{n-1} (x-Q)\Pi_{n-1}\right)^{-1}=\Pi_{n-1}\left(1 - {\bfp \bfe_n^t\over \psi_n} \right) \Pi_n R \Pi_{n-1}
\eeq

\item
$\bullet$ inverse of $\Pi_{n-1}{1\over x-Q}\Pi_{n-1}$:
\beq
 \left(\Pi_{n-1}{1\over x-Q}\Pi_{n-1}\right)^{-1}
= \Pi_{n-1} (1 + {\bfe_{n-1}\hat\bfq^t\over \hat\phi_{n-1}} \Pi^{n})(x-Q)\Pi_{n-1}
\eeq

\item
$\bullet$ inverse of $\Pi_{n-1} (x-Q)(y-P^t) \Pi_{n-1}$:
\beq\label{invPix-Qy-PPi}
 (\Pi_{n-1} (x-Q)(y-P^t) \Pi_{n-1})^{-1} = \Pi_{n-1} \td{R}^t \Pi_n \left(1-{1\over K_{n+1}}\bfp \bfq^t\right) \Pi_{n} R\Pi_{n-1}
\eeq

\item
$\bullet$ inverse of $\Pi_{n-1} {1\over x-Q}{1\over y-P^t} \Pi_{n-1}$:

\bea
&& (\Pi_{n-1} {1\over x-Q}{1\over y-P^t} \Pi_{n-1})^{-1} \cr
&=& \Pi_{n-1}(y-P^t)(\Pi_{n-2}+{1\over J_{n-1}}(1-\Pi_{n-2})\hat\bfp\hat\bfq^t(1-\Pi_{n-2}))(x-Q)\Pi_{n-1} \cr
\eea

\item
$\bullet$ equivalent formula for $\Pi_{n-1} {1\over x-Q}{1\over y-P^t} \Pi_{n-1}$:

\bea\label{invQPJQPiP}
&&  \Pi_{n-1} {1\over x-Q}{1\over y-P^t} \Pi_{n-1} \cr
&=&  \Pi_{n-1} {1\over x-Q}\Pi_{n-1}{1\over y-P^t} \Pi_{n-1}+J_n\Pi_{n-1}\bfp\bfq^t\Pi_{n-1} \cr
&=&  \Pi_{n-1} {1\over x-Q}\Pi_{n-2}{1\over y-P^t} \Pi_{n-1}+J_{n-1}\Pi_{n-1}\bfp\bfq^t\Pi_{n-1} \cr
\eea

\end{enumerate}

\section{Windows}

The Christoffel Darboux theorems of section \ref{secCDth}, show that the kernels can be computed in terms of only $\psi_m$ with $n-d_2\leq m \leq n$ and
$\hat\phi_m$ with $n-1\leq m\leq n+d_2-1$ (resp. $\phi_m$ with $n-d_1\leq m \leq n$ and $\hat\psi_m$ with $n-1\leq m\leq n+d_1-1$).
We thus introduce the following vectors, called ``windows'' \cite{BEHduality, BEHRH, BEformulaD}:
\beq
\wPsi_n(x) = (\psi_{n-d_2},\psi_{n-d_2+1},\dots,\psi_{n-1},\psi_n)^t = \Pi_n^{n-d_2}\bfp
\eeq
and similarly:
\beq
\wPhi_n(y) = \Pi_n^{n-d_1}\bfq
\virg
\hat\wPsi_n(y) = \Pi_{n+d_1-1}^{n-1}\hat\bfp
\virg
\hat\wPhi_n(y) = \Pi_{n+d_2-1}^{n-1}\hat\bfq
\eeq

We also introduce the following matrices, called ``windows'' \cite{BEHduality, BEHRH, BEformulaD}:

\beq
\Psi_n(x)_{mi}:=\psi_m^{(i)}(x) \virg m=n-d_2,\dots,n \virg i=0,\dots,d_2
\eeq
\beq
\Phi_n(y)_{mi}:=\phi_m^{(i)}(y) \virg m=n-d_1,\dots,n \virg i=0,\dots,d_1
\eeq
\beq
\hat\Psi_n(y)_{mi}:=\hat\psi_m^{(i)}(y) \virg m=n-1,\dots,n+d_1-1 \virg i=0,\dots,d_1
\eeq
\beq
\hat\Phi_n(x)_{mi}:=\hat\phi_m^{(i)}(x) \virg m=n-1,\dots,n+d_2-1 \virg i=0,\dots,d_2
\eeq

The relationship \eq{hatKx1egalx2} implies the duality (found in \cite{BEHduality}):
\beq\label{eqduality}
\hat\Phi_n^t(x) A_n \Psi_n(x) = \Id_{d_2+1}
\eeq
and similarly:
\beq
\Phi_n^t(y) B_n^t \hat\Psi_n(y) = \Id_{d_1+1}
\eeq

Notice that (combining \eq{relL-RPsiPhi} and \eq{eqduality}):
\beq\label{Psifolded}
\Psi_\infty(x) = (L(x)-R(x))A_n \Psi_n(x)
\eeq

The matrix
\beq\label{defFolding}
F_n(x):=(L(x)-R(x))A_n
\eeq
is the so called Folding matrix of \cite{BEformulaD}.
It's  property is to fold any operator acting on $\Psi_\infty$ into an operator acting on the window only.
For any finite band operator ${\hat{O}}$, we have:
\beq
\Pi_{n}^{n-d_2}\,{\hat{O}} \Psi_\infty(x) = \left( \Pi_{n}^{n-d_2}\, {\hat{O}}\, F_n(x) \right)\,\Psi_n(x)
\eeq
$\Pi_{n}^{n-d_2}\, {\hat{O}}\, F_n(x)$ is a square matrix of size $d_2+1\times d_2+1$. When acting on the window $\Psi_n$, it gives the same result as the operator ${\hat{O}}$ acting on the infinite matrix $\Psi_\infty$.

\section{Mixed correlation function}
\label{secmixedfction}

We now arrive to the main results of this article, which concerns the computation of the mixed correlation function
\beq\label{defWn}
W_n(x,y)=1+\left< \Tr{1\over x-M_1}{1\over y-M_2}\right>,
\eeq
where $M_1$ and $M_2$ are in the ensemble $H_n\times H_n(\Gamma)$ of remark \ref{remGammamatrixmodel}, and explained in \cite{eynhabilit},
with the measure:
\beq
{1\over Z}\,\ee{-\Tr(V_1(M_1)+V_2(M_2)+M_1 M_2)} \,\, dM_1\,dM_2.
\eeq
and where $Z$ is  the normalization constant, called partition function:
\beq
Z =\int_{H_n\times H_n(\Gamma)} \ee{-\Tr(V_1(M_1)+V_2(M_2)+M_1 M_2)} \,\, dM_1\,dM_2.
\eeq
We recall that $H_n\times H_n(\Gamma)$ is the set of normal matrices (i.e. $[M_1,M_1^\dagger]=0=[M_2,M_2^\dagger]$) with pairs of eigenvalues constrained to be on $\Gamma$, see \cite{eynhabilit} for more details.
In  the case $\Gamma=\R\times \R$, $M_1$ and $M_2$ are hermitean matrices of size $n$.

\medskip

The formula of \cite{BEmixed} reads:
\bea\label{Wndetntimesn}
W_{n}(x,y)
&=& \mathop{\det}_{n}{\left( \Id_{n} + \Pi_{n-1} {1\over x-Q}{1\over y-P^t}\Pi_{n-1} \right)} \cr
\eea
which is a $n\times n$ determinant, and is thus not convenient for large $n$ computations, and for many other applications.

The purpose of this section is to write it in terms of a determinant of size $d_2+1$ or $d_1+1$.

\subsection{Formula for $W_n$}

We introduce the following lower triangular matrices\footnote{They are infinite matrices, with only a  non-vanishing sub-block of size  $d_1+1$ (resp. $d_2+1$).
} of size $d_1+1$ (resp. $d_2+1$):
\bea\label{defU}
U_n(x,y) &:=& -{y+V'_1(x)\over \gamma_n}\bfe_n\bfe_{n-1}^t - \Pi_n {V'_1(x)-V'_1(Q)\over x-Q} \Pi^{n-1} \cr
\tdU_n(x,y) &:=& -{x+V'_2(y)\over \gamma_n}\bfe_n\bfe_{n-1}^t - \Pi_n {V'_2(y)-V'_2(P)\over y-P} \Pi^{n-1} \cr
\eea
Many of their properties are described in appendix C.

The following theorem is proved in appendix D:

\bt\label{thWn}
\bea\label{formulaWnsingledet}
&& W_n(x,y) \cr
&=& \gamma_n^2\,K_{n+1}(x,y)\,J_{n-1}(x,y)\,\det(\Id_{n}-\Pi_n(1-{\bfp\bfq^t\over K_{n+1}})U_n(1-{\hat\bfp\hat\bfq^t\over J_{n-1}})\tdU_n^t) \cr
&=& \gamma_n^2\,K_{n+1}(x,y)\,J_{n-1}(x,y)\,\det(\Id_{d_2+1}-\Pi_n^{n-d_2}(1-{\wPsi_n\wPhi_n^t\over K_{n+1}})U_n(1-{\hat\wPsi_n\hat\wPhi_n^t\over J_{n-1}})\tdU_n^t) \cr
&=& \gamma_n^2\,K_{n+1}(x,y)\,J_{n-1}(x,y)\,\det(\Id_{d_1+1}-\Pi_n^{n-d_1}(1-{\hat\wPsi_n\hat\wPhi_n^t\over J_{n-1}})\tdU_n^t (1-{\wPsi_n\wPhi_n^t\over K_{n+1}})U_n) \cr
\eea
or also:
\beq\label{mainformulaWn}
\encadremath{
\begin{array}{lll}
W_n(x,y)
&=& \gamma_n^2\, \det(\Id_{n+1}-U_n(x,y)\tdU_n(x,y)^t) \cr
&& \left( (J_{n+d_2}+\hat\wPhi_n^t {1\over 1- \tdU_n^t U_n}  \hat\wPsi_n)\,(K_{n-d_2}+\wPhi_n^t{1\over 1-U_n\tdU_n^t}\wPsi_n) \right. \cr
&& \qquad \left. - (\hat\wPhi_n^t \tdU_n^t {1\over 1-U_n\tdU_n^t} \wPsi_n)\,(\wPhi_n^t{1\over 1-U_n\tdU_n^t} U_n \hat\wPsi_n) \right) \cr
\end{array}}
\eeq

\et
Notice that those formulae involve only functions which are within the windows.
The determinants are in fact of size min$(d_2+1,d_1+1)$.

\medskip
The proof is given in appendix D.

\bt\label{thWnkernels}
\beq
W_n(x,y) = -\td{t}\,K_{n-d_2}(x,y)\,J_{n+d_2}(x,y)\,\det (y-M(x,y))
\eeq
where $M$ is the $(d_2+1)\times (d_2+1)$ matrix
\beq
M_{ij}=H_n^{(ij)}(x,x)+{\delta_{i0}\over K_{n-d_2}(x,y)}(y-\partial_x)K_{n-d_2}^{(0j)}(x,y)+{\delta_{j0}\over J_{n+d_2}(x,y)}(y+\partial_x)J_{n+d_2}^{(i0)}(x,y)
\eeq

\et

\proof{We use expressions of appendix F, and Lemma \ref{lemmausefulformula}.}

\subsection{Recursion $W_{n+1}-W_{n}$}

We introduce the following lower triangular matrices of size $d_1$ (resp. $d_2$):
\beq\label{defcalW}
\calW_n(x) = \Pi_n{V'_1(x)-V'_1(Q)\over x-Q}\Pi^n
\virg
\td\calW_n(y) = \Pi_n{V'_2(y)-V'_2(P)\over y-P}\Pi^n
\eeq

The following theorem is proved in appendix E:

\bt\label{threcWn}
\bea
&&  W_{n+1}(x,y)-W_{n}(x,y) \cr
&=& K_{n+1}J_n\,\det(1-\Pi_n(1-{\bfp\bfq^t\over K_{n+1}})\calW_n (1-{\hat\bfp\hat\bfq^t\over J_n}) \td\calW_n^t) \cr
\eea
i.e.
\beq
\encadremath{
\begin{array}{lll}
&& W_{n+1}(x,y)-W_{n}(x,y) \cr
&=& \det(1-\calW_n\td\calW_n^t)\,\,\, .  \quad [(J_n+(\hat\bfq^t \calW_n^t{1\over 1-\calW_n\td\calW_n^t}\td\calW_n\hat\bfp))(\bfq^t {1\over 1-\calW_n\td\calW_n^t}\bfp) \cr
&& \qquad -(\bfq^t {1\over 1-\calW_n\td\calW_n^t}\calW_n\hat\bfp)(\hat\bfq^t \td\calW_n^t{1\over 1-\calW_n\td\calW_n^t}\bfp)] \cr
\end{array}
}\eeq
\et

i.e. $W_{n+1}-W_n$ actualy involves the computation of a determinant and inverse of a matrix of size ${\rm min}(d_1,d_2)$.

\medskip
The proof is given in appendix E.

\section{Application: differential systems and spectral curve}
\label{secdifsys}

An important observation of \cite{BEHduality} and \cite{BEHRH, KML, kapaev} is that windows of consecutive bi-orthogonal polynomials satisfy some integrable differential systems,
and a Riemann--Hilbert problem. An explicit representation of those differential systems was found in \cite{BEformulaD}.
Here, thanks to the result for the mixed correlation function, we are able to give new representations of those systems, and compute explicitely their spectral curve.

\subsection{Differential systems}

We define the following $(d_2+1)\times (d_2+1)$ and $(d_1+1)\times (d_1+1)$ matrices:
\beq
\Dcal_n(x):=\Psi'_n(x) \,\Psi_n^{-1}(x)
\virg
\hat\Dcal_n(x):=-\hat\Phi'_n(x) \,\hat\Phi_n^{-1}(x)
\eeq
\beq
\td\Dcal_n(y):=\Phi'_n(y) \,\Phi_n^{-1}(y)
\virg
\hat{\td\Dcal}_n(y):=-\hat\Psi'_n(y) \,\hat\Psi_n^{-1}(y).
\eeq
Those matrices are square matrices of the size of the corresponding window, and they have polynomial entries.
They give some ODE's for the windows:
\beq
\Psi'_n(x) = \Dcal_n(x)\,\Psi_n(x).
\eeq

It was found in \cite{BEHduality} that they enjoy some duality relations (which merely come from duality \eq{eqduality}):
\beq
\hat\Dcal_n^t(x) A_n = A_n \Dcal_n(x)
\virg
\hat{\td\Dcal}_n^t(y) B_n = B_n \td\Dcal_n(y).
\eeq

\bt\label{thdifsysD}
We have the following equivalent expressions for ${\cal D}$:
\beq\label{eqthdifsysD}
\encadremath{
\begin{array}{lll}
\Dcal_n(x) &=& \Pi_{n}^{n-d_2} P^t (L(x)-R(x)) A_n \cr
y-\Dcal_n(x) &=& (\td{U}_n^{t -1}(x,y) - \Pi_{n}^{n-d_2} U_n(x,y)) A_n \cr
\left(\Psi_n^{-1}(x)\, \Dcal_n(x)\,\Psi_n(x)\right)_{ij} &=& H^{(ij)}(x,x) \cr
\end{array}
}\eeq

And the matrix elements of $y-\Dcal_n(x)$ are $(d_1+1)\times (d_1+1)$ determinants given by ($0\leq k,l\leq d_2$):

- If $k\geq d_1$
\bea
 \left(y-\Dcal_n(x)\right)_{n-k,n-l} = {1\over \prod_{m=1}^{d_1} Q_{n-m,n-m-d_2}}\,\times \cr
 \det\pmatrix{
\bfe_{n-k}^t (y-P^t)\Pi_{n-d_2-1}^{n-d_1-d_2} & &\bfe_{n-k}^t (y-P^t)\bfe_{n-l}\cr
& & \cr
\Pi_{n-1}^{n-d_1} (x-Q)\Pi_{n-d_2-1}^{n-d_1-d_2} & &\Pi_{n-1}^{n-d_1} (x-Q)\bfe_{n-l}\cr
} \cr
\eea

- If $k< d_1$
\bea
 \left(y-\Dcal_n(x)\right)_{n-k,n-l} = {(-1)^{d_1-k}\over \prod_{m=0}^{d_1-k} \gamma_{n+m-1}\, \prod_{m=1}^{k} Q_{n-m,n-m-d_2}}\, \times \det \cr
  \pmatrix{
\bfe_{n-k}^t (y-P^t)\Pi^{n-d_2-k}_{n-d_2-1} & \!\!\bfe_{n-k}^t (y-P^t)\bfe_{n-l} &  \!\!\bfe_{n-k}^t (y-P^t)\Pi_{n+d_1-k}^{n+1} \cr
& &  \cr
\Pi_{n+d_1-k-1}^{n-k} (x-Q)\Pi^{n-d_2-k}_{n-d_2-1} & \!\!\Pi_{n+d_1-k-1}^{n-k} (x-Q)\bfe_{n-l} &  \!\!\Pi_{n+d_1-k-1}^{n-k} (x-Q)\Pi_{n+d_1-k}^{n+1} \cr
} \cr
\eea

\et

The first equality of \eq{eqthdifsysD} was found in \cite{BEformulaD}, it comes from the Folding matrix \eq{defFolding}: $\Psi'_n = \Pi_n^{n-d_2} \Psi_\infty' = \Pi_n^{n-d_2} P^t \Psi_\infty = \Pi_n^{n-d_2} P^t F_n(x) \Psi_n$.

The second equality is a rewriting of the first one using the matrices $U_n$ and $\td{U}_n$, see appendix C.

The third equality is a mere rewriting of the definition of ${\cal D}$, using the duality \eq{hatKx1egalx2}.

The last representation is obtained by using the first formula, and computing the folding matrix by inverting the linear problem $\Psi_\infty = F_n \Psi_n$ in the kernel of $x-Q$.

\subsection{Spectral curve}

It was proven in \cite{BEHduality} that all those differential systems share the same spectral curve.
We define:
\bea
\Ecal_n(x,y) &:=&  \td{t}\,\det(y-\Dcal_n(x))= \td{t}\,\det(y-\hat\Dcal_n(x)) \cr
                  &=&  t\,\det(x-\td\Dcal_n(y))= t\,\det(x-\td{\hat\Dcal}_n(y)) \cr
\eea
We are going to give here several equivalent formulae for computing $\Ecal$.

\bt\label{thmarcorec}
We have:
\beq
\encadremath{
\Ecal_n(x,y)  = -\gamma_n^2\, \det(1-U_n\td{U}_n^t)
}\eeq
and
\beq
\Ecal_{n+1}(x,y)-\Ecal_n(x,y)  = - \det(1-\calW_n\td{\calW}_n^t)
\eeq

\et
\proof{For the first expression, use the 2nd expression in theorem \ref{thdifsysD}. The proof of the recursion formula is found in appendix H.}

\medskip

Then, we introduce the following Lemma, which consists in taking the polynomial part at large $y$ of formula \eq{mainformulaWn}:
\bl\label{thformulaUdetH}
\beq
\encadremath{
 \mathop{{\rm Pol}}_{y\to\infty} (V'_2(y)+x) W_n(x,y)
= \td{t} \det_{i,j\neq 0}(y-H^{(ij)})
}\eeq
i.e.
\beq
V'_2(y)+x+\left<\Tr{1\over x-M_1}\,{V'_2(y)-V'_2(M_2)\over y-M_2}\right> = \td{t} \det_{i,j\neq 0}(y-H^{(ij)})
\eeq
\el
Notice that this formula is the finite $n$ counterpart of what was found in the formal large $n$ expansion in \cite{CEO2mat}, namely theorem 3.1, equation (3.3) of \cite{CEO2mat}.

\proof{This Lemma is proved in appendix F.5.}

\medskip

Using this Lemma, taking the polynomial part in $x$ at large $x$, we prove the following theorem, which was conjectured by Marco Bertola \cite{marcoconj}:

\bt\label{thmarco}
Proof of Bertola's conjecture:
\beq
 \mathop{{\rm Pol}}_{x\to\infty}   \mathop{{\rm Pol}}_{y\to\infty} (V'_1(x)+y)(V'_2(y)+x) W_n(x,y)  = 2n + \Ecal_n(x,y)
\eeq
i.e.
\beq
\encadremath{
\Ecal_n(x,y) = (V'_1(x)+y)(V'_2(y)+x)+\left<\Tr{V'_1(x)-V'_1(M_1)\over x-M_1}\,{V'_2(y)-V'_2(M_2)\over y-M_2}\right>-n
}\eeq
\et
Marco Bertola has proved it for potentials of degree max$(d_1,d_2)\leq 5$, which is quite remarkable, and also for the smallest values of $n$ \cite{marcoconj}.

This theorem is also the finite $n$ counterpart of theorem 3.1 of \cite{CEO2mat}.
Notice that the spectral curve for finite $n$ is the same as the algebraic curve found from large $n$ considerations \cite{CEO2mat}.
This fact has important consequences, some of which are described below in the next section.

\proof{The proof is in appendix G.}

\bigskip

For completeness, we also write a  formula for the spectral curve, due to Jacques Hurtubise:
\bt
Hurtubise formula \cite{hurtubisepr}
\beq
\Ecal_n(x,y) = -\,{\gamma_n\over \prod_{j=n-d_2}^{n+d_1} \gamma_j}\,\,\det\pmatrix{
\Pi_{n}^{n-d_2}\,(P^t-y)\cr \Pi_{n+d_1-1}^{n-1}\,(Q-x) }
\eeq
\et
which simply amounts to say that if $\Ecal_n(x,y)=0$, there must exist some functions $\psi_{n-d_2-1},\dots,\psi_{n+d_1}$, such that we have simultaneously $y\psi_i(x)=\psi_i'(x)=\sum_i P_{ji} \psi_j(x)$ for $i=n-d_2,\dots, n$
and $x\psi_i(x)=\sum_i Q_{ij}\psi_j(x)$ for $i=n-1,\dots,n+d_1-1$,
 i.e. the vector $(\psi_{n-d_2-1},\dots,\psi_{n+d_1})^t$ is in the kernel of the matrix above.

\section{Examples}

\subsection{Gaussian case, Ginibre polynomials}

Consider $V_1(x)=V_2(y)=0$, which is Ginibre's ensemble.
It is  well known that the bi-orthogonal polynomials are monomials \cite{ginibre}:
\beq
p_n(x) = x^n
\virg
q_n(y) = y^n
\virg
h_n=  n! \pi,
\eeq
and the Cauchy transforms are
\beq
\hat{p}_n(y) = {\pi\,n!\over y^{n+1}}
\virg
\hat{q}_n(x) = {\pi\,n!\over x^{n+1}}.
\eeq
We have:
\beq
Q_{nm}=P_{nm} = \sqrt{n+1}\,\,\delta_{m,n+1}
\virg
\gamma_n=\sqrt{n}.
\eeq
We find that for $n>m$:
\beq
R_{nm}(x) = -\psi_n(x)\hat\phi_m(x),
\eeq
and ${1\over x-Q}$ is an upper triangular matrix.

In that case, theorem \ref{thWn} becomes:
\bea
W_n(x,y)
&=& 1+n J_{n-1}(x,y) K_n(x,y)  - xy J_n(x,y) K_{n-1}(x,y)  \cr
&=& n K_{n+1} J_{n-1} - xy K_n J_n,
\eea
and theorem \rf{threcWn} with $\calW_n=0$ becomes:
\beq
W_{n+1}-W_n =  K_{n+1} J_n.
\eeq

\subsection{Gaussian elliptical case}

Consider $V'_1(x)=t  x$, $V'_2(y)=\td{t} y$, we write:
\beq
\delta=1-t \td{t}
\eeq
The bi-orthogonal polynomials are rescaled Hermite polynomials \cite{Mehtabio}
%\beq
%p_n(x) = \delta^{-n/2}\,H_n(\sqrt\delta\,x)
%\virg
%q_n(x) = \delta^{-n/2}\,H_n(\sqrt\delta\,y)
%\eeq
, and we have:
\beq
\gamma_n=Q_{n-1,n}=P_{n-1,n} = \sqrt{n\over \delta}
\eeq
\beq
Q_{n,n-1} = -\td{t} \gamma_n
\virg
P_{n,n-1} = -t  \gamma_n
\eeq
The Christoffel--Darboux matrices are:
\beq
A_n= \gamma_n \pmatrix{0 & 1 \cr \td{t} & 0}
\virg
B_n= \gamma_n \pmatrix{0 & 1 \cr t  & 0}
\eeq
The matrices $U_n$, $\td{U}_n$ are:
\beq
U_n(x,y) = -\pmatrix{t  & 0 \cr {y+t  x\over \gamma_n} & t  }
\virg
\td{U}^t_n(x,y) = -\pmatrix{\td{t} &  {x+\td{t} y\over \gamma_n} \cr 0 & \td{t} }
\eeq
and the matrices ${\cal W}_n$ and $\td{\cal W}_n$ are of dimension $1$:
\beq
\calW(x) = t
\virg
\td\calW(y) = \td{t}
\eeq
%thus
%\beq
%\calM_n(x,y) = t  \td{t} {J_{n+1}(x,y)\over J_n(x,y)}\,\bfe_n\bfe_n^t
%\eeq
%Formula \eq{mainformula} gives:

The differential systems are:
\beq
{\cal D}_n(x) = \pmatrix{-{x\over \td{t}} & {\delta\over\td{t}}\gamma_n \cr \delta\gamma_n & -t  x}
\virg
\overline{\cal D}_n(x) = \pmatrix{-t  x & \delta\gamma_n \cr {\delta\over\td{t}}\gamma_n & -{x\over \td{t}}}
\eeq
\beq
\td{\cal D}_n(y) = \pmatrix{-{y\over t } & {\delta\over t }\gamma_n \cr \delta\gamma_n & -\td{t} y}
\virg
\overline{\td{\cal D}}_n(y) = \pmatrix{-\td{t} y & \delta\gamma_n \cr {\delta\over t }\gamma_n & -{y\over t }}
\eeq
thus the spectral curve is:
\beq
{\cal E}_n(x,y) = \td{t} \det(y-{\cal D}_n(x)) = (x+\td{t} y)(y+t  x) - \delta n = -\gamma_n^2 \det(1-U_n\td{U}_n^t)
\eeq

Theorem \ref{threcWn} gives:
\beq
W_{n+1}-W_n = J_n K_{n+1} - t  \td{t} J_{n+1} K_n
\eeq

\section{Tau-function}

Theorem \ref{thmarco} implies for $k\leq d_1$ and $l\leq d_2$:
\beq
\left<\Tr M_1^k\, M_2^l\right> = \Res_{x\to\infty}\,\Res_{y\to\infty}\,{\Ecal_n(x,y)\, x^k\,y^l\over V'_1(x)\,V'_2(y)}\,dx \,dy
\eeq
Thus, if we introduce the matrix integral:
\beq
\td{Z}=\int dM_1\, dM_2\,\,\ee{-[\sum_{k,l} t_{k,l}\,\Tr M_1^k M_2^l]}
\eeq
we have:
\beq\label{eqdZresEcal}
\Res_{x\to\infty}\,\Res_{y\to\infty}\,{\Ecal_n(x,y)\, x^k\,y^l\over V'_1(x)\,V'_2(y)}\,dx \,dy = -\left. {\d \ln{\td{Z}}\over \d t_{k,l}}\right|_{t_{k,l}=0\,\,{\rm if}\, kl>1}
\eeq
Equation \eq{eqdZresEcal} is similar to the tau-function of \cite{JM, JMU, UT, FIK} generalized to the resonant case by \cite{bertolatau}, and thus, similarly to \cite{BEHiso}, we have proved that:
\beq
\tau(\{ t_{k,l} \}) = \td{Z} \qquad {\rm for}\,t_{k,l}=0\,\,{\rm if}\, kl>1.
\eeq
This strongly suggests that the general setting of the isomonodromic problem should include all $t_{k,l}$'s.

It is easy to define bi-orthogonal polynomials with a weight of the form
$$
\om(x,y) = \ee{-\sum_{k,l} t_{k,l} x^k y^l}
$$
and they are related to the so called ensemble of normal complex matrix integrals (see \cite{WZ}):
\beq
Z_{\hbox{normal}} = \int_{N_n} dM\, \ee{-\sum_{k,l} t_{k,l} \Tr M^k M^{\dagger l}}
\eeq
where $N_n$ is the set of complex matrices which commute with their adjoint $[M,M^\dagger]=0$ (they have complex eigenvalues not constrained on any path).
For normal matrices, we have:
\beq
\left<\Tr {1\over x-M}{1\over y-M^\dagger}\right> = \Tr \left(\Pi_{n-1}{1\over y-P^t}{1\over x-Q}\Pi_{n-1}\right)
\eeq
so that our $W_n$ is  not the mixed correlation function for normal matrices.

Our $W_n(x,y)$ is in that model:
\beq
W_n(x,y) = \left<\det\left(1+{1\over x-M}{1\over y-M^\dagger}\right)\right>=\det(1+\Pi_{n-1}{1\over x-Q}{1\over y-P^t}\Pi_{n-1}).
\eeq
It is not clear yet why this function should be related to the spectral curve for the normal matrix model...

\section{Conclusion}

In this article, we have found a formula for computing the mixed correlation function as a ${\rm min}(d_1+1,d_2+1)$ determinant, instead of a $n\times n$ determinant.
This kind of formula can be very useful for finding large $n$ limits.
Beside, we have found several expressions for the spectral curve, which can be useful for studying the integrability properties.
In particular, we have proved the conjecture of Marco Bertola, which should open the route to other applications.

\medskip
In this article we have computed the 2-point mixed correlation function, however, for many applications to physics (BMN limit of ADS/CFT correspondance in string theory, or boundary conformal field theory),
one needs mixed correlation functions involving more than one trace, and more than two matrices in a trace.
The key element of the work of \cite{BEmixed}, was the use of Morozov's formula for 2-point correlation functions of unitary integrals. Since then, some generalizations of Morozov's formula have been found for arbitrary correlation functions \cite{eynprats}, and it seems natural to generalize the result of \cite{BEmixed}, and of the present paper.
In that prospect, one could expect to understand the Bethe-ansatz-like structure of large $n$ mixed correlation, which was found in \cite{OEmixed}.

\subsection*{Aknowledgements}

We would like to thank Marco Bertola, John Harnad, Jacques Hurtubise and Ken Mc Laughlin for fruitful discussions on those subjects.
Our work is partly supported by the Enigma european network MRT-CT-2004-5652, by the ANR project G\'eom\'etrie et int\'egrabilit\'e en physique math\'ematique ANR-05-BLAN-0029-01,
and by the Enrage european network MRTN-CT-2004-005616.
B.E. thanks the CRM and the BIRS for their supports.

%-----------------------------------------------------------------------------------------------------------------

\vfill\eject
\setcounter{section}0
\setcounter{equation}0

\appendix{Proof of theorem \ref{thleftinv}}
\label{proofthleftinv}

\proof{
Notice that $\hat\psi_n(y)$ is piecewise analytical in the connected domains separated by the $\td\gamma^{(i)}$'s. In each such domain, it bahaves at $\infty$, like:
\beq
\hat\psi_n(y)\sim_\infty \sqrt{h_n}\,y^{-n-1}\,\ee{V_2(y)}\, (1+O(1/y))
\eeq
Let $f_n(x,y)$ be the primitive of $\hat\psi_n(y)\ee{xy}$ which vanishes at $\infty$ in sectors where $\Re V_2<0$:
\beq
f_n(x,y) = \int^y \hat\psi_n(y'')\ee{xy''}
\eeq
$f_n(x,y)$ behaves at $\infty$ in each connected domain, like:
\beq
f_n(x,y)\sim_\infty {1\over \td{t}}\,\sqrt{h_n}\,y^{-n-d_2-1}\,\ee{V_2(y)+xy}\,(1+\sum_{k=1}^\infty f_{n,k}(x)\,y^{-k})
\eeq
where each $f_{n,k}(x)$ is a polynomial in $x$.

By definition of $L_{nm}(x)$ we have:
\bea
L_{nm}(x)
&=& {1\over 2i\pi}\,\sum_{i=1}^{d_2}\,\int_{\td{\gamma}^{(i)}} dy\, \phi_m(y)\,\ee{-xy}\, (f_n(x,y_+)-f_n(x,y_-)) \cr
&=& {1\over 2i\pi}\,\sum_{{\rm domains}\, D}\,\int_{\partial D} dy\, \phi_m(y)\,\ee{-xy}\, f_n(x,y) \cr
&=& {1\over 2i\pi}\,\sum_{{\rm domains}\, D}\,\int_{\partial D} dy\, q_m(y)\,y^{-n-d_2-1}\,\ee{-xy}\, (1+\sum_{k=1}^\infty f_{n,k}(x)\,y^{-k}) \cr
&=& \Res_{\infty} dy\, q_m(y)\,y^{-n-d_2-1}\,\ee{-xy}\, (1+\sum_{k=1}^\infty f_{n,k}(x)\,y^{-k}) \cr
\eea
This expression vanishes if $m<n+d_2$, and is clearly a polynomial in $x$.

Now, compute:
\bea
&& \sum_{i=1}^{d_2} \psi_n^{(i)}(x) \hat\phi_m^{(i)}(x) \cr
&=& {1\over 2i\pi\,\sqrt{h_n h_m}}\,\sum_{i=1}^{d_2} \int_{\td\gamma^{(i)}} dy \int_{\tilde{\overline{\gamma}}^{(i)}} dy'' \int_\Gamma dx' dy'\,\, \ee{-V_2(y)} \ee{V_2(y'')} \ee{-V_2(y')} \ee{-V_1(x')} \cr
&&{1\over x-x'} {V'_2(y'')-V'_2(y')\over y''-y'} \, p_n(x') q_m(y) \,\,  \ee{xy''}\ee{-xy}\ee{-x'  y'} \cr
&=& \sum_{i=1}^{d_2}\sum_{j,k}{\kappa_{kj}\over 2i\pi\,\sqrt{h_n h_m}}\, \int_{\td\gamma^{(i)}} dy \int_{\tilde{\overline{\gamma}}^{(i)}} dy'' \int_{\gamma^{(k)}} dx'  \int_{\td\gamma^{(j)}} dy'\,\, \ee{-V_2(y)} \ee{V_2(y'')} \ee{-V_2(y')} \ee{-V_1(x')} \cr
&&{1\over x-x'} {V'_2(y'')-V'_2(y')\over y''-y'} \, p_n(x') q_m(y) \,\,  \ee{xy''}\ee{-xy}\ee{-x'  y'} \cr
&=& \sum_{i=1}^{d_2}\sum_{j,k}{\kappa_{kj}\over 2i\pi\,\sqrt{h_n h_m}}\, \int_{\td\gamma^{(i)}} dy \int_{\tilde{\overline{\gamma}}_+^{(i)}(y)\cup\tilde{\overline{\gamma}}_-^{(i)}(y) } dy'' \int_{\gamma^{(k)}} dx'  \int_{\td\gamma^{(j)}} dy'\,\, \cr
&& \ee{-V_2(y)} \ee{V_2(y'')} \ee{-V_2(y')} \ee{-V_1(x')} \, {1\over x-x'} {V'_2(y'')-V'_2(y')\over y''-y'} \, p_n(x') q_m(y) \,\,  \ee{xy''}\ee{-xy}\ee{-x'  y'} \cr
\eea

Moreover we have:
\bea
&=&  {1\over 2i\pi}\,\int_{\tilde{\overline{\gamma}}_+^{(i)}(y)\cup\tilde{\overline{\gamma}}_-^{(i)}(y) } dy''  \int_{\td\gamma^{(j)}} dy'\,\,  \ee{V_2(y'')} \ee{-V_2(y')}\, {V'_2(y'')-V'_2(y')\over y''-y'} \,  \ee{xy''} \ee{-x'  y'} \cr
&=& {1\over 2i\pi}\,\int_{\tilde{\overline{\gamma}}_+^{(i)}(y)\cup\tilde{\overline{\gamma}}_-^{(i)}(y) } dy'' \int_{\tilde{{\gamma}}^{(j)}}\, dy'\, {V'_2(y'')\over y''-y'} \ee{V_2(y'')} \ee{-V_2(y')}\,\ee{-x' y'}\, \ee{xy''}  \cr
&& -{1\over 2i\pi}\,\int_{\tilde{\overline{\gamma}}_+^{(i)}(y)\cup\tilde{\overline{\gamma}}_-^{(i)}(y) } dy'' \int_{\tilde{{\gamma}}^{(j)}}\, dy'\, {V'_2(y')\over y''-y'} \ee{V_2(y'')} \ee{-V_2(y')}\,\ee{-x' y'}\, \ee{xy''}  \cr
&=&  {1\over 2i\pi}\,\int_{\tilde{\overline{\gamma}}_+^{(i)}(y)\cup\tilde{\overline{\gamma}}_-^{(i)}(y) } dy'' \int_{\tilde{{\gamma}}^{(j)}}\, dy'\, {1\over (y''-y')^2} \ee{V_2(y'')} \ee{-V_2(y')}\,\ee{-x' y'}\, \ee{xy''}  \cr
&& - {1\over 2i\pi}\,\int_{\tilde{\overline{\gamma}}_+^{(i)}(y)\cup\tilde{\overline{\gamma}}_-^{(i)}(y) } dy'' \int_{\tilde{{\gamma}}^{(j)}}\, dy'\, {x\over y''-y'} \ee{V_2(y'')} \ee{-V_2(y')}\,\ee{-x' y'}\, \ee{xy''}  \cr
&& - {\delta_{ij}\over 2i\pi}\, \int_{\tilde{{\gamma}}^{(j)}}\, dy'\, \left[{1\over y''-y'} \ee{V_2(y'')} \ee{-V_2(y')}\,\ee{-x' y'}\, \ee{xy''}\right]  \cr
&& - {1\over 2i\pi}\,\int_{\tilde{\overline{\gamma}}_+^{(i)}(y)\cup\tilde{\overline{\gamma}}_-^{(i)}(y) } dy'' \int_{\tilde{{\gamma}}^{(j)}}\, dy'\, {1\over (y''-y')^2} \ee{V_2(y'')} \ee{-V_2(y')}\,\ee{-x' y'}\, \ee{xy''}  \cr
&& + {1\over 2i\pi}\,\int_{\tilde{\overline{\gamma}}_+^{(i)}(y)\cup\tilde{\overline{\gamma}}_-^{(i)}(y) } dy'' \int_{\tilde{{\gamma}}^{(j)}}\, dy'\, {x'\over y''-y'} \ee{V_2(y'')} \ee{-V_2(y')}\,\ee{-x' y'}\, \ee{xy''}  \cr
&=&  \delta_{ij}\, \Res_{y} dy'\, {1\over y-y'} \ee{V_2(y)} \ee{-V_2(y')}\,\ee{-x' y'}\, \ee{x y}  \cr
&& - {1\over 2i\pi}\,\int_{\tilde{\overline{\gamma}}_+^{(i)}(y)\cup\tilde{\overline{\gamma}}_-^{(i)}(y) } dy'' \int_{\tilde{{\gamma}}^{(j)}}\, dy'\, {x-x'\over y''-y'} \ee{V_2(y'')} \ee{-V_2(y')}\,\ee{-x' y'}\, \ee{xy''}  \cr
&=&  - \delta_{ij}\,  \ee{(x-x') y}  \cr
&& - {1\over 2i\pi}\,\int_{\tilde{\overline{\gamma}}_+^{(i)}(y)\cup\tilde{\overline{\gamma}}_-^{(i)}(y) } dy'' \int_{\tilde{{\gamma}}^{(j)}}\, dy'\, {x-x'\over y''-y'} \ee{V_2(y'')} \ee{-V_2(y')}\,\ee{-x' y'}\, \ee{xy''}  \cr
\eea

Therefore:
\bea
&& \sum_{i=1}^{d_2} \psi_n^{(i)}(x) \hat\phi_m^{(i)}(x) \cr
&=& - \sum_{i=1}^{d_2}\sum_{j,k}{\kappa_{kj}\delta_{ij}\over \sqrt{h_n h_m}}\, \int_{\td\gamma^{(i)}} dy  \int_{\gamma^{(k)}} dx'  \,\, \ee{-V_2(y)}  \ee{-V_1(x')}\, {1\over x-x'}  \, p_n(x') q_m(y) \,\,  \ee{-x' y} \cr
&& - \sum_{i=1}^{d_2}\sum_{j,k}{\kappa_{kj}\over 2i\pi\,\sqrt{h_n h_m}}\, \int_{\td\gamma^{(i)}} dy \int_{\tilde{\overline{\gamma}}_+^{(i)}(y)\cup\tilde{\overline{\gamma}}_-^{(i)}(y) } dy'' \int_{\gamma^{(k)}} dx'  \int_{\td\gamma^{(j)}} dy'\,\, \cr
&& \qquad \ee{-V_2(y)} \ee{V_2(y'')} \ee{-V_2(y')} \ee{-V_1(x')}\, {1\over y''-y'} \, p_n(x') q_m(y) \,\,  \ee{xy''}\ee{-xy}\ee{-x'  y'} \cr
&=& - \left({1\over x-Q}\right)_{nm} \cr
&& - \sum_{i=1}^{d_2}\sum_{j,k}{\kappa_{kj}\over 2i\pi\,\sqrt{h_n h_m}}\, \int_{\td\gamma^{(i)}} dy \int_{\tilde{\overline{\gamma}}_+^{(i)}(y)\cup\tilde{\overline{\gamma}}_-^{(i)}(y) } dy'' \int_{\gamma^{(k)}} dx'  \int_{\td\gamma^{(j)}} dy'\,\, \cr
&& \qquad \ee{-V_2(y)} \ee{V_2(y'')} \ee{-V_2(y')} \ee{-V_1(x')} \, {1\over y''-y'} \, p_n(x') q_m(y) \,\,  \ee{xy''}\ee{-xy}\ee{-x'  y'} \cr
&=& - \left({1\over x-Q}\right)_{nm}  - \sum_{i=1}^{d_2}{1\over 2i\pi}\, \int_{\td\gamma^{(i)}} dy \int_{\tilde{\overline{\gamma}}_+^{(i)}(y)\cup\tilde{\overline{\gamma}}_-^{(i)}(y) } dy''\, \hat\psi_n(y'') \phi_m(y) \,\,  \ee{xy''}\ee{-xy} \cr
&=& - \left({1\over x-Q}\right)_{nm} + L_{nm}(x)
\eea
Using \eq{recxFourrier}, this also proves that $L$ is  a left inverse of $x-Q$.

}

\appendix{Proof of determinantal formulae}
\label{appproofdetformula}

\subsection{Inverses}

\begin{enumerate}

\item
\bea
&& \left( \Pi_{n-1}\left(1 - {\bfp \bfe_n^t\over \psi_n} \right) \Pi_n R \Pi_{n-1}\right).\left(\Pi_{n-1} (x-Q)\Pi_{n-1}\right) \cr
&=& \Pi_{n-1}\left(1 - {\bfp \bfe_n^t\over \psi_n} \right) \Pi_n R (x-Q)\Pi_{n-1} \cr
&=& \Pi_{n-1}\left(1 - {\bfp \bfe_n^t\over \psi_n} \right) \Pi_n (1-{\bfp \bfe_0^t\over \psi_0})\Pi_{n-1} \cr
&=& \Pi_{n-1}\left(1 - {\bfp \bfe_n^t\over \psi_n} -{\bfp \bfe_0^t\over \psi_0}+{\bfp \bfe_n^t\over \psi_n}{\bfp \bfe_0^t\over \psi_0}\right)\Pi_{n-1} \cr
&=& \Pi_{n-1}\left(1 - {\bfp \bfe_n^t\over \psi_n} \right)\Pi_{n-1} \cr
&=& \Id_n
\eea

\item inverse of $\Pi_{n-1}{1\over x-Q}\Pi_{n-1}$:
\bea
&& \left(\Pi_{n-1}{1\over x-Q}\Pi_{n-1}\right) .\left( \Pi_{n-1} (1 + {\bfe_{n-1}\hat\bfq^t\over \hat\phi_{n-1}} \Pi^{n})(x-Q)\Pi_{n-1}\right) \cr
&=& \Pi_{n-1}(R+\bfp \hat\bfq^t\Pi_{n-1}) (1 + {\bfe_{n-1}\hat\bfq^t\over \hat\phi_{n-1}} \Pi^{n})(x-Q)\Pi_{n-1} \cr
&=& \Pi_{n-1}( R + R{\bfe_{n-1}\hat\bfq^t\over \hat\phi_{n-1}} \Pi^{n}+\bfp \hat\bfq^t\Pi_{n-1} + \bfp \hat\bfq^t{\bfe_{n-1}\hat\bfq^t\over \hat\phi_{n-1}} \Pi^{n})\cr
&& \qquad \qquad (x-Q)\Pi_{n-1} \cr
&=& \Pi_{n-1}( R + \bfp \hat\bfq^t(\Pi_{n-1} +  \Pi^{n}))(x-Q)\Pi_{n-1} \cr
&=& \Pi_{n-1}( R + \bfp \hat\bfq^t)(x-Q)\Pi_{n-1} \cr
&=& \Id_n
\eea

\item inverse of $\Pi_{n-1} (x-Q)(y-P^t) \Pi_{n-1}$:
\bea
& (\Pi_{n-1} (x-Q)(y-P^t) \Pi_{n-1}).\left( \Pi_{n-1} \td{R}^t \Pi_n \left(1-{1\over K_{n+1}}\bfp \bfq^t\right) \Pi_{n} R\Pi_{n-1}\right) \cr
=& \Pi_{n-1} (x-Q)(y-P^t)  \td{R}^t \Pi_n \left(1-{1\over K_{n+1}}\bfp \bfq^t\right) \Pi_{n} R\Pi_{n-1} \cr
=& \Pi_{n-1} (x-Q)(\Pi_n-{\bfe_0 \bfq^t\over \phi_0}\Pi_n)  \left(1-{1\over K_{n+1}}\bfp \bfq^t\right) \Pi_{n} R\Pi_{n-1} \cr
=& \Pi_{n-1} (x-Q)(\Pi_n-\Pi_n{\bfp \bfq^t\over K_{n+1}}-{\bfe_0 \bfq^t\over \phi_0}\Pi_n+{\bfe_0 \bfq^t\over \phi_0}\Pi_n{\bfp \bfq^t\over K_{n+1}})   \Pi_{n} R\Pi_{n-1} \cr
=& \Pi_{n-1} (x-Q)(\Pi_n-\Pi_n{\bfp \bfq^t\over K_{n+1}}-{\bfe_0 \bfq^t\over \phi_0}\Pi_n+{\bfe_0 \bfq^t\over \phi_0})   \Pi_{n} R\Pi_{n-1} \cr
=& \Pi_{n-1} (x-Q)(\Pi_n-\Pi_n{\bfp \bfq^t\over K_{n+1}}\Pi_n)  R\Pi_{n-1} \cr
=& \Pi_{n-1} (x-Q)(1-{\bfp \bfq^t\over K_{n+1}}\Pi_n)  R\Pi_{n-1} \cr
=& \Pi_{n-1} (x-Q)  R\Pi_{n-1} \cr
=& \Id_n
\eea

\item
$\bullet$
Proof that:
\bea
&&  \Pi_{n-1} {1\over x-Q}{1\over y-P^t} \Pi_{n-1} \cr
&=&  \Pi_{n-1} {1\over x-Q}\Pi_{n-1}{1\over y-P^t} \Pi_{n-1}+J_n\Pi_{n-1}\bfp\bfq^t\Pi_{n-1} \cr
&=&  \Pi_{n-1} {1\over x-Q}\Pi_{n-2}{1\over y-P^t} \Pi_{n-1}+J_{n-1}\Pi_{n-1}\bfp\bfq^t\Pi_{n-1} \cr
\eea

We have:
\bea
&& \left({1\over x-Q}{1\over y-P^t}\right)_{nm} \cr
&=& {1\over \sqrt{h_n\,h_m}}\,\int_\Gamma q_m(y') {1\over y-y'}\,{1\over x-x'} p_n(x') \,\,\om(x',y')\,dx'\, dy'  \cr
&=& \!\!\!\!{1\over \sqrt{h_n\,h_m}}\,\int_\Gamma  {q_m(y)-(q_m(y)-q_m(y'))\over y-y'}\,{p_n(x)-(p_n(x)-p_n(x'))\over x-x'}  d\mu(x',y')  \cr
&=& {1\over \sqrt{h_n\,h_m}}\,\int_\Gamma  {q_m(y)\over y-y'}\,{p_n(x)\over x-x'}  \,\,\om(x',y')\,dx'\, dy'  \cr
&& -{1\over \sqrt{h_n\,h_m}}\,\int_\Gamma  {q_m(y)-q_m(y')\over y-y'}\,{p_n(x)\over x-x'}  \,\,\om(x',y')\,dx'\, dy'  \cr
&& -{1\over \sqrt{h_n\,h_m}}\,\int_\Gamma  {q_m(y)\over y-y'}\,{p_n(x)-p_n(x')\over x-x'}  \,\,\om(x',y')\,dx'\, dy'  \cr
&& +{1\over \sqrt{h_n\,h_m}}\,\int_\Gamma  {q_m(y)-q_m(y')\over y-y'}\,{p_n(x)-p_n(x')\over x-x'}  \,\,\om(x',y')\,dx'\, dy'  \cr
&=& \psi_n(x)\phi_m(y)\,J_0(x,y) \cr
&& + \sum_{l=0}^{m-1} {1\over \sqrt{h_n\,h_m}}\,\int_\Gamma  \td{R}_{ml}(y) q_l(y')\,{p_n(x)\over x-x'}  \,\,\om(x',y')\,dx'\, dy'  \cr
&& + \sum_{j=0}^{n-1} {1\over \sqrt{h_n\,h_m}}\,\int_\Gamma  {q_m(y)\over y-y'}\,R_{nj}(x) p_j(x')  \,\,\om(x',y')\,dx'\, dy'  \cr
&& +\sum_{j=0}^{n-1}\sum_{l=0}^{m-1} {1\over \sqrt{h_n\,h_m}}\,\int_\Gamma  \td{R}_{ml}(y)\,R_{nj}(x)\,q_l(y') p_j(x')\,\om(x',y')\,dx'\, dy'  \cr
&=& \Big( J_0(x,y)\,\bfp(x)\bfq^t(y)  + \bfp(x) \hat\bfq^t(x)\td{R}^t(y) +\cr
&& \qquad \quad  R(x) \hat\bfp(y) \bfq^t(y)  + R(x)\td{R}^t(y) \Big)_{nm}
\eea
Thus:
\beq
{1\over x-Q}{1\over y-P^t} =  J_0\,\bfp\bfq^t  + \bfp \hat\bfq^t\td{R}^t + R \hat\bfp \bfq^t  + R\td{R}^t
\eeq
Now, we take the $\Pi_{n-1}$ projection:
\bea
&& \Pi_{n-1}{1\over x-Q}{1\over y-P^t}\Pi_{n-1} - J_0\, \Pi_{n-1}\bfp\bfq^t\Pi_{n-1}\cr
&=&   \Pi_{n-1}\bfp \hat\bfq^t\td{R}^t\Pi_{n-1} + \Pi_{n-1}R \hat\bfp \bfq^t\Pi_{n-1}  + \Pi_{n-1}R\td{R}^t\Pi_{n-1} \cr
&=&  \Pi_{n-1}\bfp \hat\bfq^t\Pi_{n-1}\td{R}^t\Pi_{n-1} + \Pi_{n-1}R\Pi_{n-1} \hat\bfp \bfq^t\Pi_{n-1}  + \cr
&& \qquad  \Pi_{n-1}R\Pi_{n-1}\td{R}^t\Pi_{n-1} \cr
&=&  \Pi_{n-1}(R+\bfp \hat\bfq^t)\Pi_{n-1}(\td{R}^t+\hat\bfp \bfq^t)\Pi_{n-1} \cr
&& \qquad -  \Pi_{n-1}\bfp\hat\bfq^t\Pi_{n-1} \hat\bfp \bfq^t\Pi_{n-1}   \cr
&=&  (J_n-J_0)\, \Pi_{n-1}\bfp\bfq^t\Pi_{n-1}  \cr
&& \qquad + \Pi_{n-1}(R+\bfp \hat\bfq^t)\Pi_{n-1}(\td{R}^t+\hat\bfp \bfq^t)\Pi_{n-1} \cr
&=&  \Pi_{n-1} {1\over x-Q}\Pi_{n-1}{1\over y-P^t} \Pi_{n-1}+J_n\Pi_{n-1}\bfp\bfq^t\Pi_{n-1}
\eea
Similarly we have:
\bea
&& \Pi_{n-1}{1\over x-Q}{1\over y-P^t}\Pi_{n-1} -  J_0\, \Pi_{n-1}\bfp\bfq^t\Pi_{n-1} \cr
&=&  \Pi_{n-1}\bfp \hat\bfq^t\td{R}^t\Pi_{n-1} + \Pi_{n-1}R \hat\bfp \bfq^t\Pi_{n-1}  + \Pi_{n-1}R\td{R}^t\Pi_{n-1} \cr
&=&  \Pi_{n-1}\bfp \hat\bfq^t\Pi_{n-2}\td{R}^t\Pi_{n-1} + \Pi_{n-1}R\Pi_{n-2} \hat\bfp \bfq^t\Pi_{n-1}  + \Pi_{n-1}R\Pi_{n-2}\td{R}^t\Pi_{n-1} \cr
&=& \Pi_{n-1}(R+\bfp \hat\bfq^t)\Pi_{n-2}(\td{R}^t+\hat\bfp \bfq^t)\Pi_{n-1} -  \Pi_{n-1}\bfp\hat\bfq^t\Pi_{n-2} \hat\bfp \bfq^t\Pi_{n-1}   \cr
&=& ( J_{n-1}-J_0)\, \Pi_{n-1}\bfp\bfq^t\Pi_{n-1}  + \Pi_{n-1}(R+\bfp \hat\bfq^t)\Pi_{n-2}(\td{R}^t+\hat\bfp \bfq^t)\Pi_{n-1} \cr
&=&  \Pi_{n-1} {1\over x-Q}\Pi_{n-2}{1\over y-P^t} \Pi_{n-1}+J_{n-1}\Pi_{n-1}\bfp\bfq^t\Pi_{n-1}
\eea

\item
$\bullet$ inverse of $\Pi_{n-1} {1\over x-Q}{1\over y-P^t} \Pi_{n-1}$:

First, notice that:
\bea
\hat\bfq^t\Pi^{n-1}(x-Q)\Pi_{n-1}
&=& \hat\bfq^t (1-\Pi_{n-2})(x-Q)\Pi_{n-1} \cr
&=& \hat\bfq^t (x-Q)\Pi_{n-1} - \hat\bfq^t \Pi_{n-2}(x-Q)\Pi_{n-1}\cr
&=& {\bfe_0^t\over \psi_0} - \hat\bfq^t \Pi_{n-2}(x-Q) \cr
\eea
Now, compute:
\bea
&& \Pi_{n-1}(y-P^t)(\Pi_{n-2}+\Pi^{n-1}{\hat\bfp\hat\bfq^t\over J_{n-1}} \Pi^{n-1})(x-Q)  \Pi_{n-1}  \cr
&=& \Pi_{n-1}(y-P^t)(\Pi_{n-2}(x-Q)+{1\over J_{n-1}}\Pi^{n-1}\hat\bfp ( {\bfe_0^t\over \psi_0} - \hat\bfq^t \Pi_{n-2}(x-Q) ))  \cr
&=& \Pi_{n-1}(y-P^t)((1-\Pi^{n-1}{\hat\bfp\hat\bfq^t\over J_{n-1}})\Pi_{n-2}(x-Q)+\Pi^{n-1}{\hat\bfp  \bfe_0^t\over J_{n-1}\psi_0}  )  \cr
\eea
Thus:
\bea
&& \Pi_{n-1}(y-P^t)(\Pi_{n-2}+\Pi^{n-1}{\hat\bfp\hat\bfq^t\over J_{n-1}} \Pi^{n-1})(x-Q)  \Pi_{n-1} {1\over x-Q}{1\over y-P^t}\Pi_{n-1} \cr
&=& \Pi_{n-1}(y-P^t)(1-\Pi^{n-1}{\hat\bfp\hat\bfq^t\over J_{n-1}})\Pi_{n-2}{1\over y-P^t}\Pi_{n-1} \cr
&& +\Pi_{n-1}(y-P^t)\Pi^{n-1}{\hat\bfp  \over J_{n-1}\psi_0}\bfe_0^t{1\over x-Q}{1\over y-P^t}\Pi_{n-1}    \cr
&=& \Pi_{n-1}(y-P^t)(1-\Pi^{n-1}{\hat\bfp\hat\bfq^t\over J_{n-1}})\Pi_{n-2}(\td{R}^t+\hat\bfp \bfq^t) \Pi_{n-1} \cr
&& +\Pi_{n-1}(y-P^t)\Pi^{n-1}{\hat\bfp \over J_{n-1}\psi_0}\bfe_0^t(R+\bfp \hat\bfq^t)\Pi_{n-2}(\td{R}^t+\hat\bfp \bfq^t) \Pi_{n-1}    \cr
&& +\Pi_{n-1}(y-P^t)\Pi^{n-1}{\hat\bfp \over J_{n-1}\psi_0}\bfe_0^t (J_{n-1} \bfp\bfq^t) \Pi_{n-1}    \cr
&=& \Pi_{n-1}(y-P^t)(1-\Pi^{n-1}{\hat\bfp\hat\bfq^t\over J_{n-1}})(\td{R}^t+\Pi_{n-2}\hat\bfp \bfq^t) \Pi_{n-1} \cr
&& +\Pi_{n-1}(y-P^t)\Pi^{n-1}{\hat\bfp \over J_{n-1}} \hat\bfq^t (\td{R}^t+\Pi_{n-2}\hat\bfp \bfq^t) \Pi_{n-1}    \cr
&& +\Pi_{n-1}(y-P^t)\Pi^{n-1}\hat\bfp \bfq^t \Pi_{n-1}    \cr
&=& \Pi_{n-1}(y-P^t)(\td{R}^t+\Pi_{n-2}\hat\bfp \bfq^t) \Pi_{n-1}  +\Pi_{n-1}(y-P^t)\Pi^{n-1}\hat\bfp \bfq^t \Pi_{n-1}    \cr
&=& \Pi_{n-1}(y-P^t)(\td{R}^t+\hat\bfp \bfq^t) \Pi_{n-1}     \cr
&=& \Id_n
\eea

\end{enumerate}

\subsection{Determinants}

\begin{enumerate}

\item
$\bullet$ determinant of $\Pi_{n-1}(x-Q)\Pi_{n-1}$.
This is a classical result and can be found in the litterature.

One possible proof, is that $\det(\Pi_{n-1}(x-Q)\Pi_{n-1})$ is a monic polynomial of degree $n$.
Consider $x$ a zeroe of $p_n(x)$, this implies that $\Pi_{n-1}\bfp(x)=\Pi_{n}\bfp(x)$, thus:
\beq
\Pi_{n-1}(x-Q)\Pi_{n-1}\bfp(x) =\Pi_{n-1}(x-Q)\Pi_{n}\bfp(x) = \Pi_{n-1}(x-Q)\bfp(x)=0
\eeq
Thus, the $n$ zeroes of $\psi_n$ are also the zeroes of $\det(\Pi_{n-1}(x-Q)\Pi_{n-1})$. The converse is easy too.

Thus we have:
\beq
\det(\Pi_{n-1}(x-Q)\Pi_{n-1}) =  p_n(x)
\virg
\det(\Pi_{n-1}(y-P)\Pi_{n-1}) =  q_n(y)
\eeq

\item
$\bullet$ determinant of $\Pi_{n-1}{1\over x-Q}\Pi_{n-1}$:

Compute:
\bea
&& p_n(x)\,\det(\Pi_{n-1}{1\over x-Q}\Pi_{n-1}) \cr
&=& \det(\Pi_{n-1}(x-Q)\Pi_{n-1}{1\over x-Q}\Pi_{n-1}) \cr
&=& \det(\Pi_{n-1}(x-Q)\Pi_{n-1}(R+\bfp \hat\bfq^t)\Pi_{n-1}) \cr
&=& \det(\Pi_{n-1}(x-Q)(\Pi_{n}-\bfe_n\bfe_n^t)(R+\bfp \hat\bfq^t)\Pi_{n-1}) \cr
&=& \det(\Pi_{n-1}(x-Q)(R+\bfp \hat\bfq^t)\Pi_{n-1} \cr
&& \qquad \quad -\Pi_{n-1}(x-Q)\bfe_n\bfe_n^t(R+\bfp \hat\bfq^t)\Pi_{n-1}) \cr
&=& \det(\Id_n + \gamma_n \bfe_{n-1}\bfe_n^t(R+\bfp \hat\bfq^t)\Pi_{n-1}) \cr
&=& 1 + \gamma_n \bfe_n^t(R+\bfp \hat\bfq^t)\bfe_{n-1} \cr
&=& 1 + \gamma_n (R_{n,n-1}+\psi_n \hat\phi_{n-1}) \cr
&=& 1 + \gamma_n (-{1\over \gamma_n}+\psi_n \hat\phi_{n-1}) \cr
&=&  \gamma_n \psi_n \hat\phi_{n-1}
\eea
i.e.
\beq
 \det(\Pi_{n-1}{1\over x-Q}\Pi_{n-1})
=  {\gamma_n\,\ee{-V_1}\over \sqrt{h_n}}\,\,  \hat\phi_{n-1}
=  {\ee{-V_1} \over \sqrt{h_{n-1}}}\,\,  \hat\phi_{n-1}
\eeq

\item
$\bullet$ kernel $K_n$:
\bea
&& \det(\Pi_{n-1} (x-Q)(y-P^t) \Pi_{n-1}) \cr
&=& \det(\Pi_{n-1} (x-Q)\Pi_{n-1}(y-P^t) \Pi_{n-1} +\Pi_{n-1} (x-Q) \Pi^n (y-P^t) \Pi_{n-1}) \cr
&=& \det(\Pi_{n-1} (x-Q)\Pi_{n-1}(y-P^t) \Pi_{n-1}+\gamma_n^2 \bfe_{n-1} \bfe_{n-1}^t) \cr
&=& p_n(x)q_n(y)\,\det(\Id_n + \cr
&& \qquad + \gamma_n^2 \Pi_{n-1}\left(1 - {\bfp \bfe_n^t\over \psi_n} \right) \Pi_n R  \bfe_{n-1} \bfe_{n-1}^t \td{R}^t\Pi_n \left(1 - { \bfe_n \bfq^t\over \phi_n} \right) \Pi_{n-1} ) \cr
&=& p_n(x)q_n(y)\,\det(\Id_n+\Pi_{n-1}\left(1 - {\bfp \bfe_n^t\over \psi_n} \right) \bfe_{n} \bfe_{n}^t  \left(1 - { \bfe_n \bfq^t\over \phi_n} \right) \Pi_{n-1} ) \cr
&=& p_n(x)q_n(y)\,(1+ \bfe_{n}^t  \left(1 - { \bfe_n \bfq^t\over \phi_n} \right) \Pi_{n-1}\left(1 - {\bfp \bfe_n^t\over \psi_n} \right) \bfe_{n}  ) \cr
&=& p_n(x)q_n(y)\,(1+ {K_n\over \psi_n \phi_n}  ) \cr
&=& p_n(x)q_n(y)\,{K_{n+1}\over \psi_n \phi_n}   \cr
&=& h_n\,\ee{V_1(x)+V_2(y)}\,K_{n+1}
\eea

\item
$\bullet$ kernel $J_n$:
\bea
&& \det(\Pi_{n-1}{1\over x-Q}{1\over y-P^t}\Pi_{n-1}) \cr
&=& \det(J_n\Pi_{n-1}\bfp\bfq^t\Pi_{n-1}+\Pi_{n-1}{1\over x-Q}\Pi_{n-1}{1\over y-P^t}\Pi_{n-1}) \cr
&=& \det(\Pi_{n-1}{1\over x-Q}\Pi_{n-1}{1\over y-P^t}\Pi_{n-1}) \cr
&& (1+J_n \bfq^t(\Pi_{n-1}{1\over x-Q}\Pi_{n-1}{1\over y-P^t}\Pi_{n-1})^{-1}\bfp) \cr
&=& {\ee{-V_1-V_2}\over h_{n-1}} \hat\psi_{n-1} \hat\phi_{n-1}\cr
&& (1+J_n \bfq^t(\Pi_{n-1}{1\over y-P^t}\Pi_{n-1})^{-1}(\Pi_{n-1}{1\over x-Q}\Pi_{n-1})^{-1}\bfp) \cr
&=& {\ee{-V_1-V_2}\over h_{n-1}} \hat\psi_{n-1} \hat\phi_{n-1} (1+J_n \bfq^t \Pi_{n-1} (y-P^t) \cr
&& \quad (1 + \Pi^{n}{\hat\bfp \bfe_{n-1}^t\over \hat\psi_{n-1}} )\Pi_{n-1} (1 + {\bfe_{n-1}\hat\bfq^t\over \hat\phi_{n-1}} \Pi^{n})(x-Q)\Pi_{n-1}\bfp) \cr
&=& {\ee{-V_1-V_2}\over h_{n-1}} \hat\psi_{n-1} \hat\phi_{n-1} (1+J_n \bfq^t \Pi_{n-1} (y-P^t) \cr
&& \quad (1 + \Pi^{n}{\hat\bfp \bfe_{n-1}^t\over \hat\psi_{n-1}} )\Pi_{n-1} (1 + {\bfe_{n-1}\hat\bfq^t\over \hat\phi_{n-1}} \Pi^{n})(x-Q)\Pi_{n-1}\bfp) \cr
&=& {\ee{-V_1-V_2}\over h_{n-1}} \hat\psi_{n-1} \hat\phi_{n-1} (1+J_n \bfq^t \Pi_{n-1} (y-P^t)\cr
&& \quad (1 + \Pi^{n}{\hat\bfp \bfe_{n-1}^t\over \hat\psi_{n-1}} )(\Pi_{n-2}+\bfe_{n-1}\bfe_{n-1}^t) (1 + {\bfe_{n-1}\hat\bfq^t\over \hat\phi_{n-1}} \Pi^{n})(x-Q)\Pi_{n-1}\bfp) \cr
&=& {\ee{-V_1-V_2}\over h_{n-1}} \hat\psi_{n-1} \hat\phi_{n-1} (1+J_n \bfq^t \Pi_{n-1} (y-P^t)\cr
&& \quad (1 + \Pi^{n}{\hat\bfp \bfe_{n-1}^t\over \hat\psi_{n-1}} )\bfe_{n-1}\bfe_{n-1}^t (1 + {\bfe_{n-1}\hat\bfq^t\over \hat\phi_{n-1}} \Pi^{n})(x-Q)\Pi_{n-1}\bfp) \cr
&=& {\ee{-V_1-V_2}\over h_{n-1}} \hat\psi_{n-1} \hat\phi_{n-1} (1+J_n \bfq^t \Pi_{n-1} (y-P^t)\cr
&& \quad (\bfe_{n-1} + \Pi^{n}{\hat\bfp \over \hat\psi_{n-1}} ) (\bfe_{n-1}^t + {\hat\bfq^t\over \hat\phi_{n-1}} \Pi^{n})(x-Q)\Pi_{n-1}\bfp) \cr
&=& {\ee{-V_1-V_2}\over h_{n-1}} \hat\psi_{n-1} \hat\phi_{n-1} (1 + \cr
&& \qquad \quad +J_n \bfq^t \Pi_{n-1} (y-P^t)  \Pi^{n-1}{\hat\bfp \over \hat\psi_{n-1}}   {\hat\bfq^t\over \hat\phi_{n-1}} \Pi^{n-1}(x-Q)\Pi_{n-1}\bfp) \cr
&=& {\ee{-V_1-V_2}\over h_{n-1}} \hat\psi_{n-1} \hat\phi_{n-1} (1+J_n \bfq^t \Pi_{n-1} (y-P^t) {\hat\bfp \over \hat\psi_{n-1}}   {\hat\bfq^t\over \hat\phi_{n-1}} (x-Q)\Pi_{n-1}\bfp) \cr
&=& {\ee{-V_1-V_2}\over h_{n-1}} \hat\psi_{n-1} \hat\phi_{n-1} (1+J_n \bfq^t  {\bfe_0 \over \hat\psi_{n-1}\phi_0}   {\bfe_0^t\over \hat\phi_{n-1}\psi_0} \bfp) \cr
&=& {\ee{-V_1-V_2}\over h_{n-1}} \hat\psi_{n-1} \hat\phi_{n-1} (1+{ J_n \over \hat\psi_{n-1} \hat\phi_{n-1}} ) \cr
&=& {\ee{-V_1-V_2}\over h_{n-1}}   J_{n-1}
\eea

\end{enumerate}

\appendix{Matrices $U$ and $\td{U}$}
\label{apUntdUn}

\subsection{Definitions}

The following matrices play an important role.

Define:
\beq\label{defUbis}
U_n(x,y):=\Pi_n (y-P^t) R(x) \Pi^{n-1}
\virg
\tdU_n(x,y):=\Pi_n (x-Q^t) \td{R}(y) \Pi^{n-1}
\eeq
Notice that since $R$ is lower triangular and $P$ is finite band, $U_n(x,y)$ (resp. $\tdU_n(x,y)$) is a small lower triangular matrix of size $d_1+1$ (resp. $d_2+1$):
\beq
U_n(x,y) = \Pi_n^{n-d_1} (y-P^t) R(x) \Pi_{n+d_1-1}^{n-1}
\eeq
$U_n(x,y)$ (resp. $\tdU_n(x,y)$) is linear in $y$ (resp. $x$) and of degree at most $d_1$ in $x$ (resp. $d_2$ in $y$).

$U_n(x,y)$ can also be rewritten as:
\beq
U_n(x,y)= -{y+V'_1(x)\over \gamma_n}\bfe_n\bfe_{n-1}^t - \Pi_n {V'_1(x)-V'_1(Q)\over x-Q} \Pi^{n-1}
\eeq
\proof{
\bea
U_n(x,y)
&=& \Pi_n^{n-d_1} (y-P^t) R(x) \Pi_{n+d_1-1}^{n-1} \cr
&=& \Pi_n^{n-d_1} (y+V'_1(x)-V'_1(x)+V'_1(Q)-V'_1(Q)-P^t) R(x) \Pi_{n+d_1-1}^{n-1} \cr
&=& \Pi_n^{n-d_1} (y+V'_1(x)) R(x) \Pi_{n+d_1-1}^{n-1} - \Pi_n^{n-d_1} (V'_1(x)-V'_1(Q)) R(x) \Pi_{n+d_1-1}^{n-1} \cr
&& \qquad - \Pi_n^{n-d_1} (V'_1(Q)+P^t) R(x) \Pi_{n+d_1-1}^{n-1} \cr
&=& \Pi_n^{n-d_1} (y+V'_1(x)) R(x) \Pi_{n+d_1-1}^{n-1} - \Pi_n^{n-d_1} (V'_1(x)-V'_1(Q)) R(x) \Pi_{n+d_1-1}^{n-1} \cr
&=& (y+V'_1(x)) \Pi_n^{n-d_1}  R(x) \Pi_{n+d_1-1}^{n-1} \cr
&& \quad - \Pi_n^{n-d_1} {V'_1(x)-V'_1(Q)\over x-Q} (x-Q) R(x) \Pi_{n+d_1-1}^{n-1}  \cr
&=& -{y+V'_1(x)\over \gamma_n}\bfe_n\bfe_{n-1}^t - \Pi^{n-d_1}_n {V'_1(x)-V'_1(Q)\over x-Q} \Pi^{n-1}_{n+d_1-1} \cr
&=& -{y+V'_1(x)\over \gamma_n}\bfe_n\bfe_{n-1}^t - \Pi_n {V'_1(x)-V'_1(Q)\over x-Q} \Pi^{n-1}
\eea}

Notice also that if $n>d_2 d_1$, we have:
\beq
U_n(x,y):=\Pi_n R(x) (y-P^t)  \Pi^{n-1}
\eeq

\subsection{Some properties}

\begin{itemize}

\item Multiplication by the Christoffel-Darboux matrix:
\beq
- U_n(x,y) A_n = B_{n+1}^t+(y-P^t)\bfe_n\bfe_n^t - \Pi_n(y-P^t)R\Pi^n(x-Q)
\eeq
\beq
 - A_n U_n(x,y) = B^t_{n-1} +\bfe_{n-1}\bfe_{n-1}^t(y-P^t)  - (x-Q)\Pi_{n-1}(y-P^t)R\Pi^{n-1}
\eeq
and subsequentely:
\beq
\Phi_\infty^t U_n(x,y) A_n \Psi_\infty
= - \Phi_\infty^t B^t_{n+1} \Psi_\infty
\eeq
\beq
 \hat\Phi_\infty^t A_n U_n(x,y) \hat\Psi_\infty = - \hat\Phi_\infty^t B^t_{n-1}\hat\Psi_\infty
\eeq

\proof{
\bea
&& - U_n(x,y) A_n \cr
&=& \Pi_n(y-P^t)R(x-Q)\Pi^n - \Pi_n(y-P^t)R\Pi^n(x-Q) \cr
&=& \Pi_n(y-P^t)\Pi^n - \Pi_n(y-P^t)R\Pi^n(x-Q) \cr
&=& [\Pi_n,y-P^t]\Pi^n+(y-P^t)\bfe_n\bfe_n^t - \Pi_n(y-P^t)R\Pi^n(x-Q) \cr
&=& B_{n+1}^t+(y-P^t)\bfe_n\bfe_n^t - \Pi_n(y-P^t)R\Pi^n(x-Q) \cr
\eea

\bea
&& - A_n U_n(x,y) \cr
&=& \Pi_{n-1}(x-Q)(y-P^t)R\Pi^{n-1} - (x-Q)\Pi_{n-1}(y-P^t)R\Pi^{n-1}  \cr
&=& \Pi_{n-1}(1+(y-P^t)(x-Q))R\Pi^{n-1} - (x-Q)\Pi_{n-1}(y-P^t)R\Pi^{n-1}  \cr
&=& \Pi_{n-1}(y-P^t)\Pi^{n-1} + \Pi_{n-1}R\Pi^{n-1} - (x-Q)\Pi_{n-1}(y-P^t)R\Pi^{n-1}  \cr
&=& \Pi_{n-1}(y-P^t)\Pi^{n-1}  - (x-Q)\Pi_{n-1}(y-P^t)R\Pi^{n-1}  \cr
&=& \Pi_{n-1}[y-P^t,\Pi^{n-1}] +\Pi_{n-1}\Pi^{n-1}(y-P^t) \cr
&& \quad - (x-Q)\Pi_{n-1}(y-P^t)R\Pi^{n-1}  \cr
&=& \Pi_{n-1}[P^t,\Pi_{n-2}] +\bfe_{n-1}\bfe_{n-1}^t(y-P^t)  - (x-Q)\Pi_{n-1}(y-P^t)R\Pi^{n-1}  \cr
&=& B^t_{n-1} +\bfe_{n-1}\bfe_{n-1}^t(y-P^t)  - (x-Q)\Pi_{n-1}(y-P^t)R\Pi^{n-1}  \cr
\eea
}

\item Inverse:
\beq
\td{U}_n^{t -1}(x,y) = \Pi_n^{n-d_2} (y-P^t) L(x) \Pi_{n+d_2-1}^{n-1}
\eeq

\proof{
\bea
&& \td{U}_n^{t}(x,y) \Pi_n^{n-d_2} (y-P^t) L(x) \Pi_{n+d_2-1}^{n-1} \cr
&=& \Pi^{n-1} \td{R}^t(y) (x-Q) \Pi_n^{n-d_2} (y-P^t) L(x) \Pi_{n+d_2-1}^{n-1} \cr
&=& \Pi^{n-1} \td{R}^t(y) \Pi^n (x-Q) \Pi_n (y-P^t) \Pi_{n-1} L(x) \Pi_{n+d_2-1}^{n-1} \cr
&=& \Pi^{n-1} \td{R}^t(y) \Pi^n (x-Q)  (y-P^t) \Pi_{n-1} L(x) \Pi_{n+d_2-1}^{n-1} \cr
&=& \Pi^{n-1} \td{R}^t(y) \Pi^n  (y-P^t) (x-Q) \Pi_{n-1} L(x) \Pi_{n+d_2-1}^{n-1} \cr
&=& \Pi^{n-1} \td{R}^t(y) (y-P^t) (x-Q) L(x) \Pi_{n+d_2-1}^{n-1} \cr
&=& \Pi_{n+d_2-1}^{n-1}
\eea
}

\item Multiplication of the inverse by the Christoffel-Darboux matrix:
\beq
 - \td{U}_n^{t -1}(x,y) A_n =   B_{n-d_2}^t \Pi^{n} + (y-P^t)\bfe_n\bfe_n^t - \Pi_n^{n-d_2} (y-P^t) L(x) \Pi^{n}(x-Q)
\eeq
\beq
A_n \td{U}_n^{t -1}(x,y) =  -\Pi^{n}B^t_{n+d_2}+\Pi^{n}_{n+d_2-1}(y-P^t) -(x-Q)\Pi^{n}(y-P^t)L(x)\Pi_{n+d_2-1}^{n-1}
\eeq

In particular:
\beq
\Phi_\infty^t \td{U}_n^{t -1}(x,y) A_n \Psi_\infty
= - \Phi_\infty^t B^t_{n-d_2} \Pi_{n-1} \Psi_\infty
\eeq
and
\beq
\hat\Phi_\infty^t A_n \td{U}_n^{t -1}(x,y) \hat\wPsi = - \hat\Phi_\infty^t \Pi^{n} B^t_{n+d_2} \hat\wPsi
\eeq

\proof{
\bea
&& \td{U}_n^{t -1}(x,y) A_n \cr
&=& \Pi_n^{n-d_2} (y-P^t) L(x) [x-Q,\Pi_{n-1}] \cr
&=& \Pi_n^{n-d_2} (y-P^t) L(x) (x-Q)\Pi_{n-1} -  \Pi_n^{n-d_2} (y-P^t) L(x)\Pi_{n-1} (x-Q) \cr
&=& \Pi_n^{n-d_2} (y-P^t) \Pi_{n-1} +\gamma_{n-d_2}L_{n-d_2-1,n-1} \bfe_{n-d_2}\bfe_{n-1}^t (x-Q) \cr
\eea

\bea
&& - \td{U}_n^{t -1}(x,y) A_n \cr
&=& \Pi_n^{n-d_2} (y-P^t) L(x) [x-Q,\Pi^{n}] \cr
&=& \Pi_n^{n-d_2} (y-P^t) L(x) (x-Q)\Pi^{n} - \Pi_n^{n-d_2} (y-P^t) L(x) \Pi^{n}(x-Q) \cr
&=& \Pi_n^{n-d_2} (y-P^t) \Pi^{n} - \Pi_n^{n-d_2} (y-P^t) L(x) \Pi^{n}(x-Q) \cr
&=& [\Pi_n^{n-d_2},y-P^t] \Pi^{n} +(y-P^t)\Pi_n^{n-d_2} \Pi^{n} \cr
&& \quad - \Pi_n^{n-d_2} (y-P^t) L(x) \Pi^{n}(x-Q) \cr
&=& B_{n-d_2}^t \Pi^{n} + (y-P^t)\bfe_n\bfe_n^t - \Pi_n^{n-d_2} (y-P^t) L(x) \Pi^{n}(x-Q) \cr
\eea

Then:
\bea
&& A_n \td{U}_n^{t -1}(x,y) \cr
&=& -[x-Q,\Pi^{n}](y-P^t)L(x)\Pi_{n+d_2-1}^{n-1} \cr
&=&  \Pi^{n}(x-Q)(y-P^t)L(x)\Pi_{n+d_2-1}^{n-1} -(x-Q)\Pi^{n}(y-P^t)L(x)\Pi_{n+d_2-1}^{n-1} \cr
&=&  \Pi^{n}(1+(y-P^t)(x-Q))L(x)\Pi_{n+d_2-1}^{n-1} \cr
&& \quad -(x-Q)\Pi^{n}(y-P^t)L(x)\Pi_{n+d_2-1}^{n-1} \cr
&=&  \Pi^{n}(y-P^t)\Pi_{n+d_2-1}^{n-1}+\Pi^{n}L(x)\Pi_{n+d_2-1}^{n-1} \cr
&& \quad -(x-Q)\Pi^{n}(y-P^t)L(x)\Pi_{n+d_2-1}^{n-1}\cr
&=&  \Pi^{n}(y-P^t)\Pi_{n+d_2-1} -(x-Q)\Pi^{n}(y-P^t)L(x)\Pi_{n+d_2-1}^{n-1}\cr
&=&  \Pi^{n}[(y-P^t),\Pi_{n+d_2-1}]+\Pi^{n}_{n+d_2-1}(y-P^t) \cr
&& \quad -(x-Q)\Pi^{n}(y-P^t)L(x)\Pi_{n+d_2-1}^{n-1}\cr
&=&  -\Pi^{n}B^t_{n+d_2}+\Pi^{n}_{n+d_2-1}(y-P^t) -(x-Q)\Pi^{n}(y-P^t)L(x)\Pi_{n+d_2-1}^{n-1}\cr
\eea

}

\end{itemize}

\appendix{Proof of theorem \ref{thWn}}
\label{proofthWn}

\proof{

Let us introduce the following $n\times n$ matrices:
\beq
\zeta_n:=\Pi_{n-1}(x-Q)(y-P^t)\Pi_{n-1}
\eeq
\beq
\xi_n:=\Pi_{n-1}{1\over x-Q}{1\over y-P^t}\Pi_{n-1}
\eeq
\beq
\omega_n:=1_n+\Pi_{n-1}(y-P^t)\Pi_{n-2}(x-Q)\Pi_{n-1}
\eeq
and, according to formulae of section \ref{detformula}:
\beq
\det \zeta_n = h_n K_{n+1}
\virg
\zeta_n^{-1} = \Pi_{n-1} \td{R}^t \Pi_n \left(1-{1\over K_{n+1}}\bfp \bfq^t\right) \Pi_{n} R\Pi_{n-1}
\eeq
\beq
\det \xi_n = {J_{n-1}\over h_{n-1}}
\eeq
\beq
\xi_n^{-1} = \Pi_{n-1}(y-P^t)(\Pi_{n-2}+{1\over J_{n-1}}\Pi^{n-1}\hat\bfp\hat\bfq^t \Pi^{n-1})(x-Q)\Pi_{n-1}
\eeq

We have:
\bea
\omega_n
&=& 1_n+\Pi_{n-1}(y-P^t)\Pi_{n-2}(x-Q)\Pi_{n-1} \cr
&=& 1_n+\Pi_{n-1}(y-P^t)(x-Q)\Pi_{n-1}-\Pi_{n-1}(y-P^t)\Pi^{n-1}(x-Q)\Pi_{n-1} \cr
&=& \zeta_n-\Pi_{n-1}(y-P^t)\Pi^{n-1}(x-Q)\Pi_{n-1} \cr
&=& \zeta_n (1-{1\over \zeta_n}\Pi_{n-1}(y-P^t)\Pi^{n-1}(x-Q)\Pi_{n-1}) \cr
\eea

Thus:
\bea
&& \det \omega_n \cr
&=& \det \zeta_n\, \det(1-{1\over \zeta_n}\Pi_{n-1}(y-P^t)\Pi^{n-1}(x-Q)\Pi_{n-1}) \cr
&=& h_n K_{n+1}\, \det(1-\Pi_{n-1} \td{R}^t \Pi_n \left(1-{1\over K_{n+1}}\bfp \bfq^t\right) \Pi_{n} R \times \cr
&& \quad \times (y-P^t)\Pi^{n-1}(x-Q)\Pi_{n-1}) \cr
&=& h_n K_{n+1}\, \det(1-\Pi_{n-1} \td{R}^t \Pi_n \left(1-{1\over K_{n+1}}\bfp \bfq^t\right) U_n (x-Q)\Pi_{n-1}) \cr
&=& h_n K_{n+1}\, \det(1- \Pi_n \left(1-{1\over K_{n+1}}\bfp \bfq^t\right) U_n \td{U}_n^t ) \cr
\eea
and:
\bea
\det \omega_n
&=& h_n K_{n+1}\, \det(1- U_n \td{U}_n^t + {1\over K_{n+1}} \Pi_n \bfp \bfq^t  U_n \td{U}_n^t ) \cr
&=& h_n K_{n+1}\, \det(1- U_n \td{U}_n^t)\,\,(1 + {1\over K_{n+1}}  \bfq^t  U_n \td{U}_n^t {1\over 1-U_n \td{U}_n^t}\bfp ) \cr
&=& h_n \, \det(1- U_n \td{U}_n^t)\,\,  ( \bfq^t {1\over 1-U_n \td{U}_n^t}\bfp ) \cr
\eea

Inverse:
\bea
&& \omega_n^{-1} \cr
&=& (1-{1\over \zeta_n}\Pi_{n-1}(y-P^t)\Pi^{n-1}(x-Q)\Pi_{n-1})^{-1} \,\zeta_n^{-1} \cr
&=& (1-\Pi_{n-1} \td{R}^t \Pi_n \left(1-{\bfp \bfq^t\over K_{n+1}}\right) U_n (x-Q)\Pi_{n-1})^{-1} \,\zeta_n^{-1} \cr
&=& (1+\Pi_{n-1} \td{R}^t \Pi_n {1\over 1- \Pi_n \left(1-{\bfp \bfq^t\over K_{n+1}}\right) U_n \td{U}_n^t}\Pi_n \left(1-{\bfp \bfq^t\over K_{n+1}}\right) U_n \times \cr
&& \quad (x-Q)\Pi_{n-1}) \,\zeta_n^{-1} \cr
&=& \zeta_n^{-1} +\Pi_{n-1} \td{R}^t \Pi_n {1\over 1- \Pi_n \left(1-{\bfp \bfq^t\over K_{n+1}}\right) U_n \td{U}_n^t}\Pi_n \left(1-{\bfp \bfq^t\over K_{n+1}}\right) U_n\times \cr
&& \quad \times \td{U}_n^t  \left(1-{\bfp \bfq^t\over K_{n+1}}\right) \Pi_{n} R\Pi_{n-1} \cr
&=& \Pi_{n-1} \td{R}^t \Pi_n {1\over 1- \Pi_n \left(1-{\bfp \bfq^t\over K_{n+1}}\right) U_n \td{U}_n^t} \left(1-{\bfp \bfq^t\over K_{n+1}}\right) \Pi_{n} R\Pi_{n-1} \cr
\eea

Now, the formula of \cite{BEmixed} gives $W_n=\det(1+\xi_n)$, thus we compute:
\bea
&& W_n \cr
&=& \det(1+\xi_n)  \cr
&=& \det \xi_n\,\, \det(1+\xi_n^{-1})  \cr
&=& \det \xi_n\,\, \det(1+\Pi_{n-1}(y-P^t)(\Pi_{n-2}+{1\over J_{n-1}}\Pi^{n-1}\hat\bfp\hat\bfq^t \Pi^{n-1})(x-Q)\Pi_{n-1})  \cr
&=& \det \xi_n\,\, \det(\omega_n+{1\over J_{n-1}} \Pi_{n-1}(y-P^t) \Pi^{n-1}\hat\bfp\hat\bfq^t \Pi^{n-1} (x-Q)\Pi_{n-1})  \cr
&=& \det \xi_n\,\,\det \omega_n\,\, (1+{1\over J_{n-1}} \hat\bfq^t \Pi^{n-1} (x-Q)\Pi_{n-1} \omega_n^{-1} \Pi_{n-1}(y-P^t) \Pi^{n-1}\hat\bfp)  \cr
\eea
Thus:
\bea
&& h_{n-1}\, W_n \cr
&=& \det \omega_n\,\, (J_{n-1}+ \hat\bfq^t \Pi^{n-1} (x-Q)\Pi_{n-1} \omega_n^{-1}\Pi_{n-1}(y-P^t) \Pi^{n-1}\hat\bfp)  \cr
&=& \det \omega_n\,\, (J_{n-1}+ \hat\bfq^t \td{U}_n^t  {1\over 1- \Pi_n \left(1-{\bfp \bfq^t\over K_{n+1}}\right) U_n \td{U}_n^t} \left(1-{\bfp \bfq^t\over K_{n+1}}\right) U_n \hat\bfp)  \cr
\eea
i.e.
\bea
W_n=\gamma_n^2 K_{n+1}J_{n-1} \det\left(1-\Pi_n(1-{\bfp\bfq^t\over K_{n+1}}) U_n (1-{\hat\bfp\hat\bfq^t\over J_{n-1}}) \td{U}_n^t \right)
\eea
The two other formulae are obtained by Weinstein-Aronstein duality.

The lemma \ref{lemmausefulformula} gives \eq{mainformulaWn}.

}

\appendix{Proof of theorem \ref{threcWn}}
\label{proofthrecWn}

The formula of \cite{BEmixed} gives:
\bea
W_{n}(x,y)
&=& \mathop{\det}_{n}{\left( \Id_{n} + \pi_{n-1} {1\over x-Q}{1\over y-P^t}\pi_{n-1}^t \right)} \cr
\eea
Using \eq{invQPJQPiP}, it can be rewritten as:
\beq
W_{n}(x,y)= \mathop{\det}_{n}{\left( \Id_{n} + J_{n}(x,y) \pi_{n-1} \bfp \bfq^t \pi_{n-1}^t +\pi_{n-1}{1\over x-Q}\Pi_{n-1}{1\over y-P^t}\pi_{n-1}^t \right)}
\eeq
\beq
W_{n+1}(x,y) = \mathop{\det}_{n+1}{\left( \Id_{n+1} + J_n(x,y) \pi_n \bfp \bfq^t \pi_n^t +\pi_n{1\over x-Q}\Pi_{n-1}{1\over y-P^t}\pi_n^t \right)}
\eeq
The difference is thus the same determinant with substracting $1$ in the $n,n$ position:
\bea\label{diffWnminor}
&& W_{n+1}(x,y)-W_{n}(x,y) \cr
&=& \mathop{\det}_{n+1}{\left( \Pi_{n-1} + J_n \pi_n \bfp \bfq^t \pi_n^t +\pi_n{1\over x-Q}\Pi_{n-1}{1\over y-P^t}\pi_n^t \right)} \cr
\eea
For any arbitrary non-vanishing $\alpha$ and $\td\alpha$, multiply the matrix inside the determinant \eq{diffWnminor} on the left by
$\Id_{n+1}-\pi_{n}{\bfp \bfe_n^t\over \psi_n}+\alpha {\bfe_n \bfe_n^t\over \psi_n}$ (whose determinant is ${\alpha\over \psi_n}$),
and on the right by
$\Id_{n+1}-{\bfe_n \bfq^t\over \phi_n}\pi_{n}^t+\td\alpha {\bfe_n \bfe_n^t\over \phi_n}$ (whose determinant is ${\td\alpha\over \phi_n}$):
\beq
(\Id_{n+1}-\pi_{n}{\bfp \bfe_n^t\over \psi_n}+\alpha {\bfe_n \bfe_n^t\over \psi_n}) \Pi_{n-1}(\Id_{n+1}-{\bfe_n \bfq^t\over \phi_n}\pi_{n}^t+\td\alpha {\bfe_n \bfe_n^t\over \phi_n})
= \Pi_{n-1}
= \Id_{n+1}-\bfe_n\bfe_n^t
\eeq
\beq
 (\Id_{n+1}-\pi_{n}{\bfp \bfe_n^t\over \psi_n}+\alpha {\bfe_n \bfe_n^t\over \psi_n}) J_n \bfp\bfq^t (\Id_{n+1}-{\bfe_n \bfq^t\over \phi_n}\pi_{n}^t+\td\alpha {\bfe_n \bfe_n^t\over \phi_n})
= J_n \alpha\td\alpha \bfe_n \bfe_n^t
\eeq
\bea
&& (\Id_{n+1}-\pi_{n}{\bfp \bfe_n^t\over \psi_n}+\alpha {\bfe_n \bfe_n^t\over \psi_n})\pi_n{1\over x-Q}\Pi_{n-1}  \cr
&=& (\Pi_{n-1}(1-{\bfp \bfe_n^t\over \psi_n})+\alpha {\bfe_n \bfe_n^t\over \psi_n})(R+\bfp\hat\bfq^t)\Pi_{n-1}  \cr
&=& \Pi_{n-1}(1-{\bfp \bfe_n^t\over \psi_n})\Pi_n R\Pi_{n-1}+\alpha {\bfe_n \bfe_n^t\over \psi_n}R   +\alpha \bfe_n \hat\bfq^t\Pi_{n-1}  \cr
&=& \Pi_{n-1}(1-{\bfp \bfe_n^t\over \psi_n})\Pi_n R\Pi_{n-1} +\alpha \bfe_n T_n^t  \cr
\eea
where
\beq
T_n^t = {\bfe_n^t R\over  \psi_n}+ \hat\bfq^t\Pi_{n-1}
\virg
\td{T}_n = {\td{R}^t \bfe_n \over  \phi_n}+ \Pi_{n-1}\hat\bfp
\eeq

Finaly, using \eq{inversePix-QPi} we have:
\bea
&& {\alpha\td\alpha\over \psi_n\phi_n}\,(W_{n+1}(x,y)-W_{n}(x,y)) \cr
&=& \mathop{\det}_{n+1}( \Id_{n+1} + (J_n \alpha\td\alpha-1) \bfe_n \bfe_n^t + \cr
&& +((\Pi_{n-1}(x-Q)\Pi_{n-1})^{-1}+\alpha \bfe_n T_n^t) ((\Pi_{n-1}(y-P^t)\Pi_{n-1})^{-1}+\td\alpha \td{T}_n \bfe_n^t)   ) \cr
\eea
If we choose $\alpha\td\alpha= {1\over J_n}$, we have:
\bea
&& {1\over J_n\psi_n\phi_n}\,(W_{n+1}(x,y)-W_{n}(x,y)) \cr
&=& \mathop{\det}_{n+1}\Big( \Id_{n+1} +((\Pi_{n-1}(x-Q)\Pi_{n-1})^{-1}+\alpha \bfe_n T_n^t)
((\Pi_{n-1}(y-P^t)\Pi_{n-1})^{-1} \cr
&& \quad +\td\alpha \td{T}_n \bfe_n^t)   \Big) \cr
&=& \mathop{\det}_{n}\Big( \Id_{n} + ((\Pi_{n-1}(y-P^t)\Pi_{n-1})^{-1}+\td\alpha \td{T}_n \bfe_n^t)((\Pi_{n-1}(x-Q)\Pi_{n-1})^{-1} \cr
&& \quad +\alpha \bfe_n T_n^t)   \Big) \cr
&=& \mathop{\det}_{n}{\left( \Id_{n} + (\Pi_{n-1}(y-P^t)\Pi_{n-1})^{-1}(\Pi_{n-1}(x-Q)\Pi_{n-1})^{-1}+{1\over J_n} \td{T}_n T_n^t   \right)} \cr
\eea
Then , we multiply both sides by \eq{detPix-QPi}, i.e. by $\det(\Pi_{n-1}(x-Q)\Pi_{n-1})$ and $\det(\Pi_{n-1}(y-P^t)\Pi_{n-1})$:
\bea
&& {h_n \ee{V_1+V_2}\over J_n}\,(W_{n+1}(x,y)-W_{n}(x,y)) \cr
&=& \mathop{\det}_{n}\Big( \Id_{n} + \Pi_{n-1}(y-P^t)\Pi_{n-1}(x-Q)\Pi_{n-1} \cr
&& \quad +{1\over J_n} \Pi_{n-1}(y-P^t)\td{T}_n T_n^t (x-Q)\Pi_{n-1}   \Big) \cr
\eea

We have:
\bea
T_n^t (x-Q)\Pi_{n-1}
&=& {1\over  \psi_n}\bfe_n^t R (x-Q)\Pi_{n-1}+ \hat\bfq^t\Pi_{n-1}(x-Q)\Pi_{n-1} \cr
&=& {\over  \psi_n}\bfe_n^t (1-{\bfp\bfe_0^t\over \psi_0})\Pi_{n-1}+ \hat\bfq^t(x-Q)-\hat\bfq^t \Pi^{n}(x-Q)\Pi_{n-1} \cr
&=& -{\bfe_0^t\over \psi_0}+ {\bfe_0^t\over \psi_0} - \hat\bfq^t \Pi^{n}(x-Q)\Pi_{n-1} \cr
&=&  - \hat\bfq^t \Pi^{n}(x-Q)\Pi_{n-1} \cr
\eea
Notice also that Heisenberg's relation \eq{PQcannonical} implies:
\beq
\Id_{n} + \Pi_{n-1}(y-P^t)(x-Q)\Pi_{n-1}= \Pi_{n-1}(x-Q)(y-P^t)\Pi_{n-1}
\eeq
Thus:
\bea
&& {h_n \ee{V_1+V_2}\over J_n}\,(W_{n+1}(x,y)-W_{n}(x,y)) \cr
&=& \mathop{\det}_{n}\Big( \Id_n+\Pi_{n-1}(y-P^t)\Pi_{n-1}(x-Q)\Pi_{n-1} \cr
&& \quad +{1\over J_n} \Pi_{n-1}(y-P^t)\Pi^{n}\hat\bfp  \hat\bfq^t \Pi^{n}(x-Q)\Pi_{n-1}   \Big) \cr
&=& \mathop{\det}_{n} \Big( \Id_n+\Pi_{n-1}(y-P^t)(x-Q)\Pi_{n-1}-\Pi_{n-1}(y-P^t)\Pi^{n}(x-Q)\Pi_{n-1} \cr
&& \qquad +{1\over J_n} \Pi_{n-1}(y-P^t)\Pi^{n}\hat\bfp  \hat\bfq^t \Pi^{n}(x-Q)\Pi_{n-1}   \Big) \cr
&=& \mathop{\det}_{n} \Big( \Pi_{n-1}(x-Q)(y-P^t)\Pi_{n-1}- \Pi_{n-1}(y-P^t)\Pi^{n}(1-{\hat\bfp  \hat\bfq^t\over J_n}) \Pi^{n}(x-Q)\Pi_{n-1}   \Big) \cr
&&
\eea
Now, use \eq{detPix-Qy-PPi} and \eq{invPix-Qy-PPi}, that gives:
\bea
&& {1\over J_n K_{n+1}}\,(W_{n+1}(x,y)-W_{n}(x,y)) \cr
&=& \mathop{\det}_{n} \Big( \Id_n - \Pi_{n-1}(y-P^t)\Pi^{n}(1-{\hat\bfp  \hat\bfq^t\over J_n}) \Pi^{n}(x-Q)\Pi_{n-1}\td{R}^t\times \cr
&& \quad \Pi_n(1-{\bfp\bfq^t\over K_{n+1}})\Pi_n R \Pi_{n-1}   \Big) \cr
&=& \mathop{\det}_{n+1} \Big( \Id_{n+1} - \Pi_n (1-{\bfp\bfq^t\over K_{n+1}})\Pi_n  R \Pi_{n-1}(y-P^t)\Pi^{n}(1-{\hat\bfp  \hat\bfq^t\over J_n}) \times \cr
&& \quad  \Pi^{n}(x-Q)\Pi_{n-1}\td{R}^t\Pi_n   \Big) \cr
\eea
Now, because of \eq{PV'_1+}, we have:
\bea
 \Pi_n  R \Pi_{n-1}(y-P^t)\Pi^{n}
&=& \Pi_n  R V'_1(Q) \Pi^{n} \cr
&=& -\Pi_n  {V'_1(x)-V'_1(Q)\over x-Q} \Pi^{n} \cr
&=& -{\cal W}_n(x) \cr
\eea
Therefore:
\beq
 W_{n+1}(x,y)-W_{n}(x,y) = J_n K_{n+1}\,\mathop{\det}_{n+1} \Big( \Id_{n+1} - \Pi_n (1-{\bfp\bfq^t\over K_{n+1}}){\cal W}_n (1-{\hat\bfp  \hat\bfq^t\over J_n}) \td{\cal W}^t_n   \Big)
\eeq

The second part of the formula is obtained by applying Lemma \ref{lemmausefulformula}.

\appendix{Proof of theorem \ref{thformulaUdetH}}
\label{proofthformulaUdetH}

We start from theorem \ref{thWn}:
\beq
\begin{array}{lll}
W_n(x,y)
&=& \gamma_n^2\, \det(\Id_{n+1}-U_n(x,y)\tdU_n(x,y)^t) \cr
&& \left( (J_{n+d_2}+\hat\wPhi_n^t {1\over 1- \tdU_n^t U_n}  \hat\wPsi_n)\,(K_{n-d_2}+\wPhi_n^t{1\over 1-U_n\tdU_n^t}\wPsi_n) \right. \cr
&& \qquad \left. - (\hat\wPhi_n^t \tdU_n^t {1\over 1-U_n\tdU_n^t} \wPsi_n)\,(\wPhi_n^t{1\over 1-U_n\tdU_n^t} U_n \hat\wPsi_n) \right) \cr
\end{array}
\eeq

Let us compute the various  terms:

$\bullet$
\bea
&& \wPhi_n^t{1\over 1-U_n\tdU_n^t}\wPsi_n  \cr
&=& \wPhi_n^t {1\over 1-U_n\tdU_n^t} \wPsi_n   \cr
&=& \wPhi_n^t \tdU_n^{t -1} A_n \Psi_n\, {1\over (\tdU_n^{t -1} -U_n) A_n \Psi_n } \wPsi_n   \cr
&=& - \bfq^t B_{n-d_2}^t \Pi_{n-1} \Psi_\infty(x)\, {1\over y-\Psi_n^{-1}\Psi'_n } f^{(0)}   \cr
&=& - \bfq^t(y) \Pi_{n-d_2}B_{n-d_2}^t \Pi_{n-1} \Psi_\infty(x)\, {1\over y-\Psi_n^{-1}(x)\Psi'_n(x) } f^{(0)}   \cr
&=& \ee{-V_2(y)}\,O(y^{n-d_2-1})
\eea
where $f^{(0)}$ is the vector $(1,0,\dots,0)^t$ of dimension $d_2+1$.

At large $y$, this expression behaves like $\ee{-V_2(y)}\,O(y^{n-d_2-1})$.

$\bullet$
\bea
&& \hat\bfq^t \Pi_{n+d_2-1}^{n-1} {1\over 1- \tdU_n^t U_n} \Pi_{n+d_1-1}^{n-1} \hat\bfp  \cr
&=& f^{(0) t} \hat\Phi_n^t  {1\over \tdU_n^{t -1}- U_n} \tdU_n^{t -1} \hat\Psi_n  f^{(0)} \cr
&=& f^{(0) t} \hat\Phi_n^t A_n \Psi_n {1\over (\tdU_n^{t -1}- U_n)A_n \Psi_n} \tdU_n^{t -1} \hat\Psi_n  f^{(0)} \cr
&=& f^{(0) t}  {1\over y\Psi_n-\Psi_n' } \tdU_n^{t -1} \hat\Psi_n  f^{(0)} \cr
&=& f^{(0) t}  {1\over y- \Psi_n^{-1}\Psi_n' } \Psi_n^{-1} \tdU_n^{t -1} \hat\Psi_n  f^{(0)} \cr
&=& f^{(0) t}  {1\over y- \Psi_n^{-1}\Psi_n' } \hat\Phi_n^{t} A_n \tdU_n^{t -1} \hat\Psi_n  f^{(0)} \cr
&=& - f^{(0) t}  {1\over y- \Psi_n^{-1}(x)\Psi_n'(x) } \hat\Phi_\infty^t(x) \Pi^{n} B^t_{n+d_2} \hat\wPsi(y) \cr
&=& \ee{V_2(y)}\,O(y^{-n-d_2-1})
\eea
It behaves as $\ee{V_2(y)}\,O(y^{-n-d_2-1})$ at large $y$.

$\bullet$
\bea
&& \hat\bfq^t \tdU_n^t {1\over 1-U_n\tdU_n^t} \bfp \cr
&=& f^{(0) t} \hat\Phi_n^t  {1\over \tdU_n^{t -1} -U_n} \Psi_n f^{(0)} \cr
&=& f^{(0) t} \hat\Phi_n^t A_n \Psi_n   {1\over (\tdU_n^{t -1} -U_n)A_n \Psi_n} \Psi_n f^{(0)} \cr
&=& f^{(0) t} {1\over y\Psi_n - \Psi'_n} \Psi_n f^{(0)} \cr
&=& f^{(0) t} {1\over y - \Psi_n^{-1}\Psi'_n} f^{(0)} \cr
&=& \left({1\over y  -  H_n(x,x) }\right)^{(00)}
\eea

$\bullet$
Similarly, we have:
\bea
 \bfq^t {1\over 1-U_n\tdU_n^t} U_n \hat\bfp
&=& \left({1\over x  -  \td{H}_n(y,y) }\right)^{(00)},
\eea
but unfortunately, that formula is not sufficient to find the large $y$ behaviour.

Instead, we prove the following lemma:

\bl
\label{asympbehavUtUU}
\beq
(x+V'_2(y))\bfq^t{1\over 1-U_n\,\td{U}_n^t}U_n\,\hat\psi_n
= -1-{n\over \td{t}\,y^{d_2+1}} +O(y^{-d_2-2})
\eeq
and symetricaly:
\beq
(y+V'_1(x))\hat\bfq^t\td{U}_n^t\,{1\over 1-U_n\,\td{U}_n^t}\,\psi_n
= -1-{n\over {t}\,x^{d_1+1}} +O(x^{-d_1-2})
\eeq
\el

\proof{
Compute the large $y$ behaviour of the following matrix:
\bea
&& \Pi^{n-1}(\td{U}_n^t+(x+V'_2)\hat\psi_n \bfq^t)\Pi_n \cr
&=& \Pi^{n-1}(\td{R}^t(x-Q)+(x+V'_2)\hat\psi_n \bfq^t)\Pi_n \cr
&=& \Pi^{n-1}((\td{R}^t+\hat\bfp \bfq^t)(x-Q)+\hat\psi_n \bfq^t(V'_2(y)-V'_2(P^t))+\hat\psi_n \bfq^t(V'_2(P^t)+Q))\Pi_n \cr
&=& \Pi^{n-1}({1\over y-P^t}(x-Q)+\hat\psi_n \bfq^t(V'_2(P^t)+Q))\Pi_n
\eea
It is easy to see that this expression behaves like $O(y^{-1})$ at large $y$.
The leading term in $y$ is:
\bea
&& {1\over y}\Pi^{n-1}(x-Q)\Pi_n+\hat\psi_{n-1} \ee{-V_2(y)} {\d\over \d y}(\ee{V_2(y)}\phi_n(y))\bfe_{n-1}\bfe_n^t +O(y^{-2}) \cr
&=& {1\over y}\Pi^{n-1}(x-Q)\Pi_n+\sqrt{h_{n-1}}\,y^{-n}\,{n\over \sqrt{h_n}}\,y^{n-1} \,\bfe_{n-1}\bfe_n^t +O(y^{-2}) \cr
&=& {1\over y}\left(\Pi^{n-1}(x-Q)\Pi_n+{n\over \gamma_n}\,\bfe_{n-1}\bfe_n^t\right) +O(y^{-2}) \cr
\eea
and thus, using $U_n=-{y\over \gamma_n}\bfe_n\bfe_{n-1}^t+O(1)$, we have:
\bea
&& U_n\,(\td{U}_n^t+(x+V'_2)\hat\psi_n \bfq^t)\Pi_n \cr
&=& -{1\over \gamma_n}\bfe_n\bfe_{n-1}^t\left((x-Q)\Pi_n+{n\over \gamma_n}\,\bfe_{n-1}\bfe_n^t\right) +O(y^{-1}) \cr
&=& -{1\over \gamma_n}\bfe_n\,\left(\bfe_{n-1}^t(x-Q)\Pi_n+{n\over \gamma_n}\,\bfe_n^t\right) +O(y^{-1}) \cr
\eea
and  the following determinant is:
\bea
&& \det(1-U_n\,(\td{U}_n^t+(x+V'_2)\hat\psi_n \bfq^t)\Pi_n) \cr
&=& \det(1+{1\over \gamma_n}\bfe_n\,\left(\bfe_{n-1}^t(x-Q)\Pi_n+{n\over \gamma_n}\,\bfe_n^t\right) +O(y^{-1})) \cr
&=& 1+{1\over \gamma_n}\left(\bfe_{n-1}^t(x-Q)\bfe_n+{n\over \gamma_n}\right) +O(y^{-1})) \cr
&=& 1+\left({n\over \gamma_n^2}-1\right) +O(y^{-1})) \cr
&=& {n\over \gamma_n^2} +O(y^{-1}))
\eea
It follows:
\bea
&& {n\over \gamma_n^2} +O(y^{-1}))  \cr
&=& \det(1-U_n\,(\td{U}_n^t+(x+V'_2)\hat\psi_n \bfq^t)\Pi_n) \cr
&=& \det(1-U_n\,\td{U}_n^t+(x+V'_2)U_n\,\hat\psi_n \bfq^t)\Pi_n) \cr
&=& \det(1-U_n\,\td{U}_n^t)\,\det(1+(x+V'_2){1\over 1-U_n\,\td{U}_n^t}U_n\,\hat\psi_n \bfq^t)\Pi_n) \cr
&=& \det(1-U_n\,\td{U}_n^t)\,\left(1+(x+V'_2)\bfq^t{1\over 1-U_n\,\td{U}_n^t}U_n\,\hat\psi_n \right) \cr
\eea
that implies:
\bea
(x+V'_2)\bfq^t{1\over 1-U_n\,\td{U}_n^t}U_n\,\hat\psi_n
&=& -1+{n\over \gamma_n^2\,\det(1-U_n\,\td{U}_n^t)} +O(y^{-d_2-2})  \cr
&=& -1-{n\over \Ecal_n(x,y)} +O(y^{-d_2-2})  \cr
&=& -1-{n\over \td{t}\,y^{d_2+1}} +O(y^{-d_2-2})  \cr
\eea
similarly we have:
\beq
(y+V'_1(x))\hat\bfq^t\td{U}_n^t\,{1\over 1-U_n\,\td{U}_n^t}\,\psi_n
= -1-{n\over {t}\,x^{d_1+1}} +O(x^{-d_2-2})
\eeq
}

\subsection{U-conjecture, Lemma \ref{thformulaUdetH}}
\label{apUconj}

Using lemma \ref{asympbehavUtUU}, we have:
\bea
&& (x+V'_2(y))\det(1-U_n\,\td{U}_n^t)\,(\hat\bfq^t\td{U}_n^t\,{1\over 1-U_n\,\td{U}_n^t}\,\psi_n)\,(\bfq^t{1\over 1-U_n\,\td{U}_n^t}U_n\,\hat\psi_n) \cr
&=& -\det(1-U_n\,\td{U}_n^t)\,(\hat\bfq^t\td{U}_n^t\,{1\over 1-U_n\,\td{U}_n^t}\,\psi_n)-{n\over \gamma_n^2}\,(\hat\bfq^t\td{U}_n^t\,{1\over 1-U_n\,\td{U}_n^t}\,\psi_n) \cr
&& \quad +O(y^{-2})  \cr
&=& -\det(1-U_n\,\td{U}_n^t)\,\left({1\over y  -  H_n(x,x) }\right)^{(00)}-{n\over \gamma_n^2}\,\left({1\over y  -  H_n(x,x) }\right)^{(00)} \cr
&& \quad +O(y^{-2})  \cr
&=& {\td{t}\over \gamma_n^2}\,\det(y-H_n(x,x))\,\left({1\over y  -  H_n(x,x) }\right)^{(00)}-{n\over \gamma_n^2}\,\left({1\over y  -  H_n(x,x) }\right)^{(00)} \cr
&& \quad +O(y^{-2})  \cr
&=& {\td{t}\over \gamma_n^2}\,\det(y-H_n(x,x))\,\left({1\over y  -  H_n(x,x) }\right)^{(00)} +O(y^{-1})  \cr
\eea
The first term is a polynomial in $y$, and this proves theorem \ref{thformulaUdetH}.

\appendix{Proof of theorem \ref{thmarco}}
\label{proofthmarco}

Starting from lemma \ref{asympbehavUtUU}, we have:
\bea
&& \gamma_n^2\,\det(1-U_n\td{U}_n^t)\,(y+V'_1(x))\,\left(\hat\bfq^t\td{U}_n^t\,{1\over 1-U_n\,\td{U}_n^t}\,\psi_n\right) \cr
&=& \Ecal_n(x,y)\,(1+{n\over {t}\,x^{d_1+1}}) +O(x^{-1}) \cr
&=& \Ecal_n(x,y) + n +O(x^{-1}) \cr
\eea
from which theorem \ref{thmarco} follows.

\appendix{Proof of theorem \ref{thmarcorec}}
\label{proofthmarcorec}

Let $r_n(x)$ be the matrix:
\beq
r_n(x) = \Pi_{n+1}^{n-d_2+1} F_n(x) = \Pi_{n+1}^{n-d_2+1} + {1\over \gamma_{n+1}}\bfe_{n+1}\bfe_n^t(x-Q)
\eeq
By definition, it is such that:
\beq
\Psi_{n+1}(x) = r_n(x) \Psi_n(x)
\eeq
Thus:
\bea
\Dcal_{n+1}(x)
&=& \Psi_{n+1}' \Psi_{n+1}^{-1} = (r_n \Psi_n)'(r_n \Psi_n)^{-1} \cr
&=& r'_n r_n^{-1} + r_n \Dcal_n r_n^{-1} \cr
&=& r_n( r_n^{-1} r'_n + \Dcal_n ) r_n^{-1} \cr
\eea
Therefore:
\beq
\Ecal_{n+1} = \td{t}\det(y-\Dcal_n - r_n^{-1} r'_n)
\eeq
Now, notice that $r_n^{-1} r'_n = -{1\over Q_{n,n-d_2}} \bfe_{n-d_2}\bfe_n^t$, thus
\beq
\Ecal_{n+1}-\Ecal_n = (-1)^{d_2+1}\,{\td{t}\over Q_{n,n-d_2}} \det({\cal C}_n)
\eeq
where we have defined
\beq
{\cal C}_n(x,y) = \Pi_n^{n-d_2+1} (y+\Dcal_n(x)) \Pi_{n-1}^{n-d_2}
\eeq
We have:
\bea
{\cal C}_n
&=& \Pi_{n}^{n-d_2+1}(y-P^t)\Pi_{n-1}^{n-d_2} + \Pi_{n}^{n-d_2+1}{V'_1(x)-V'_1(Q)\over x-Q}(1-\Pi_{n-1})(x-Q) \Pi_{n-1}^{n-d_2} \cr
&=& \Pi_{n}^{n-d_2+1}(y-P^t)\Pi_{n-1}^{n-d_2} + \Pi_{n}^{n-d_2+1}\calW_n(x-Q) \Pi_{n-1}^{n-d_2} \cr
\eea
and:
\bea
{\cal C}_n \,\, \Pi_{n-1}^{n-d_2}\td{R}^t\Pi_n^{n-d_2+1}
&=& \Pi_n^{n-d_2+1} (1-\calW_n\td\calW^t_n)\Pi_n^{n-d_2+1}
\eea
That implies:
\bea
\det({\cal C}_n)
 \,\, \Pi_{n-1}^{n-d_2}\td{R}^t\Pi_n^{n-d_2+1}
&=& (-1)^{d_2}\gamma_n\dots\gamma_{n-d_2+1}\,\,\det(1-\calW_n\td\calW^t_n) \cr
&=& (-1)^{d_2+1}{Q_{n,n-d_2}\over \td{t}}\,\,\det(1-\calW_n\td\calW^t_n)  \cr
\eea
which proves the formula.

\appendix{A usefull formula for determinants with rank 2 matrices}
\label{appendixusefulformula}

\bl\label{lemmausefulformula}
if $M$ is an invertible matrix, and $a,b,c,d$ are arbitrary vectors, we have:

\beq\label{ausefulformula}
\det(M+a b^t+c d^t) = \det{M}\,\,\left( (1+b^t{1\over M}a)(1+d^t{1\over M}c)-  b^t{1\over M}c\,\, d^t{1\over M}a \right)
\eeq

\el

\end{document}